\documentclass[12pt]{article}
\usepackage{graphicx, amssymb, amsmath}
\usepackage{bm}
\usepackage{setspace}
\usepackage{natbib}
\usepackage{placeins}
\usepackage{colortbl,xcolor}
\usepackage{url}
\usepackage{etoolbox}
\RequirePackage[colorlinks,citecolor=blue,urlcolor=blue]{hyperref}
\usepackage{cancel,ulem}

\usepackage[margin=1in]{geometry}

\newcommand\blfootnote[1]{%
  \begingroup
  \renewcommand\thefootnote{}\footnote{#1}%
  \addtocounter{footnote}{-1}%
  \endgroup
}

\DeclareGraphicsExtensions{.pdf,.eps.gz,.eps,.epsi.gz,.epsi,.ps,.ps.gz}
\newcommand{\floor}[1]{\lfloor #1 \rfloor}

\newcommand{\dd}{\mathrm{d}}

\newcommand{\bbeta}{{\bm\beta}}

\newcommand{\btheta}{{\bm\theta}}

\def\t{{\bf t}}

\def\x{{\bf x}}

\def\Y{{\bf Y}}
\def\y{{\bf y}}

\def\1{{\bf 1}}
\def\0{{\bf 0}}

\newtheorem{theorem}{Theorem}%
\newtheorem{lemma}{Lemma}%

\begin{document}

\title{Learning Human Activity Patterns using Clustered Point Processes with Active and Inactive States}
\author{{Jingfei Zhang{\small $~^{1}$}, Biao Cai{\small $~^{2}$}, Xuening Zhu{\small $~^{3}$}, Hansheng Wang{\small $~^{4}$},}\\{Ganggang Xu{\small $~^{5}$} and Yongtao Guan{\small $~^{6}$}}
\vspace{2mm}\\
\fontsize{11}{10}\selectfont\itshape
$^{1,2,5,6}$\,Department of Management Science, University of Miami, Coral Gables, FL, U.S.A.\\ 
\fontsize{11}{10}\selectfont\itshape
$^{3}$\,School of Data Science, Fudan University, Shanghai, China \\ 
\fontsize{11}{10}\selectfont\itshape
$^{4}$\,Department of Business Statistics and Econometrics, Peking University, Beijing, China \\ 
}
\date{}
\maketitle
\blfootnote{The first two authors contributed equally to this work.}

\begin{abstract}
\noindent 

Modeling event patterns is a central task in a wide range of disciplines. In applications such as studying human activity patterns, events often arrive clustered with sporadic and long periods of inactivity. Such heterogeneity in event patterns poses challenges for existing point process models. In this article, we propose a new class of clustered point processes that alternate between active and inactive states. The proposed model is flexible, highly interpretable, and can provide useful insights into event patterns. A composite likelihood approach and a composite EM estimation procedure are developed for efficient and numerically stable parameter estimation. We study both the computational and statistical properties of the estimator including convergence, consistency, and asymptotic normality. The proposed method is applied to Donald Trump's Twitter data to investigate if and how his behaviors evolved before, during, and after the presidential campaign. Additionally, we analyze large-scale social media data from Sina Weibo and identify interesting groups of users with distinct behaviors.
\end{abstract}

\noindent{Keywords: clustered point processes; composite likelihood; composite EM algorithm; non-overlapping clusters; social media.}

\newpage

\newpage
\section{Introduction}
Recently, vast amounts of event time data collected from social media, financial trading, and online retail platforms have attracted keen research interests from various scientific communities. Analyzing such data can provide useful insights into human online activity patterns \citep{Ghose2011, Sun2017} and help to develop effective business practices such as advertisement placement and content recommendations. Temporal point process models, as a class of powerful statistical learning tools for event time data, have been extensively studied in the recent statistical and machine learning literature, see, e.g.,  
\cite{karimi2016smart, farajtabar2016multistage, farajtabar2017coevolve,hosseini2017recurrent,xiao2017joint}. While the existing work offers useful tools, it remains challenging to model event time data from human activities, due to the often complex mixture between periods of bursty event occurrences and sporadic and long periods of inactivity \citep{barabasi2005origin}. The main goal of this article is to propose a flexible and interpretable modeling framework that can adequately address such heterogeneity in event patterns.

\begin{figure}[t!]
	\centering
	\includegraphics[width=0.75\textwidth]{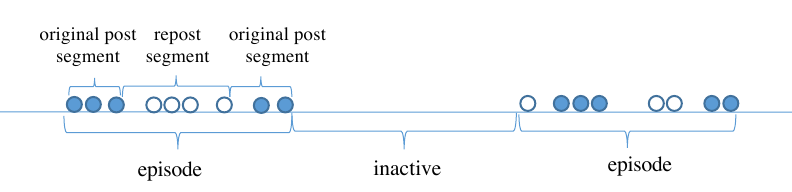}
	\caption{Illustration of episodes and segments.}
	\label{illustrate}
\end{figure}

While the proposed framework is general, we shall describe our model in the context of two motivating examples collected respectively from Twitter and Sina Weibo, a popular social media site in China. The first dataset contains tweeting times from Donald Trump (@realDonaldTrump) from January 2013 to April 2018; the second dataset contains posting times from a large sample of users of Sina Weibo in one month (see Section~\ref{sec:real} for more details). For both Twitter and Sina Weibo, a user can generate original content or repost content from other accounts. These two different types of posts are distinguished in the data and will be referred to as \textit{original posts} and \textit{reposts}, respectively. The posting times of a user are clustered, as a user's interaction with social media sites often alternate between active and inactive states \citep{Raghavan2014}. During an active state, events are generated, often with short inter-event distances; during an inactive state, no event is generated until the start of the next active state. As a result, events appear in clusters, which we subsequently refer to as \textit{episodes}. Moreover, within each episode, a user tends to publish several consecutive original posts or reposts, rendering alternating original post and repost sub-clusters. We refer to these sub-clusters as \textit{segments} in an episode (see Figure~\ref{illustrate}).

To model the above described clustered point patterns, the majority of existing approaches can be categorized into two classes: the Hawkes process \citep{Hawkes1971} and the Cox process \citep{Diggle1983}. The Hawkes process is a self-exciting process, in which the arrival of one event may trigger the occurrences of future events. It has been successfully applied in modeling information diffusion \citep{farajtabar2017coevolve}, gang activities \citep{Linderman2014}, and other scientific problems \citep{farajtabar2016multistage,zarezade2018steering, achab2018uncovering}. However, the Hawkes process may not fit well when there are long intervals of inactivity between bursts of events, which are commonly observed in human activity patterns \citep{Wu2019}. As an example, in Section~\ref{hawkes}, we demonstrate that the Hawkes process fits poorly to our motivating datasets.
The Cox process is a class of doubly stochastic point processes that include many popular models as special cases such as the log Gaussian Cox process \citep{Moller1998} and the shot noise Cox process \citep{Waagepetersen2007}. The Cox process offers limited interpretability as the formulation cannot directly characterize the active and inactive states. From a scientific point of view, we may wish to understand the transition mechanism between active and inactive states, as well as how events are generated during an active state.

In this article, we propose a new class of clustered point processes that alternate between active and inactive states. The proposed model is flexible, highly interpretable, and can provide useful insights into event patterns. The estimated parameters from different users can be used as features in other supervised and unsupervised machine learning tasks (e.g., clustering, classification); see Section~\ref{sec::sina} for an example of such an analysis. A composite likelihood approach and a composite EM estimation procedure are developed for efficient and numerically stable parameter estimation. A goodness-of-fit procedure is proposed to evaluate the model fitting. In our theoretical investigation, we establish both computational and statistical properties of the estimator including convergence, consistency, and asymptotic normality. We remark that, although motivated by social media user activity data, our method is not limited to this type of application alone. For instance, it can be applied to model transaction records of trading accounts (where an event can be either buy or sell), the arrivals of trades and quotes in financial markets \citep{engle2003trades}, and real-time smoking and alcohol usage with ecological momentary assessment \citep{cooney2009alcohol}.

The rest of the article is organized as follows. Section 2 shows results from fitting Hawkes processes to our motivating data examples. Section 3 introduces the proposed model. Section 4 describes a composite likelihood estimation approach, a composite likelihood EM algorithm, and a procedure to assess goodness of fit. Section 5 presents theoretical results, including convergence guarantee of the algorithm, as well as consistency and asymptotic normality of the estimator. Section 6 includes simulation studies. Section 7 applies the proposed method to Donald Trump's Twitter data and the Sina Weibo user data. A final discussion section concludes the article.

\section{Model fitting using bivariate Hawkes processes}

In this section, we fit stationary and nonstationary Hawkes processes to the bivariate event time data (i.e., original posts and reposts) discussed in Section 1, and illustrate the limitations of the Hawkes process in such types of applications.

Consider the observation window $[0,T]$. Let $N_i(B)$ denote the number of events in $B\subset\mathbb{R}$ for the $i$th process, $i=1,2$. The intensity functions can be defined as
\begin{equation}\label{eqn:hawkes}
\lambda_i(t)=\lim_{\Delta\downarrow0}\frac{E[N_i(t,t+\Delta)|\mathcal{H}_t]}{\Delta}, \quad i=1,2, 
\end{equation}
where $\mathcal{H}_t$ denotes the entire event time history up to time $t$. Specifically, the intensity functions of a bivariate Hawkes process take the form
\begin{equation}
\label{hawkes}
\lambda_i(t)=\phi\left[\nu_i(t)+\int_0^t\omega_{ii}(t-s) N_i(\dd s)+\int_0^t\omega_{ij}(t-s) N_j(\dd s)\right],
\end{equation}
where $\phi(\cdot)$ is a link function, $\nu_i(t)>0$ is the background intensity for the $i$th point process, and $\omega_{ii}(\cdot)$ and $\omega_{ij}(\cdot)$ are some transfer functions, for $i,j=1,2$ and $i\ne j$. 
Since the transfer functions are typically assumed to be nonnegative, past events will increase current values of the intensity functions in \eqref{hawkes}, which is commonly known as the ``self-exciting" property. If $\nu_i(t)=\nu_i$ for some constant $\nu_i>0, i=1,2$, then the resulting Hawkes process is stationary. Otherwise, it is nonstationary. 


\begin{figure}[!t]
	\centering
	\makebox{\includegraphics[trim=5mm 0 0 0, scale=0.75]{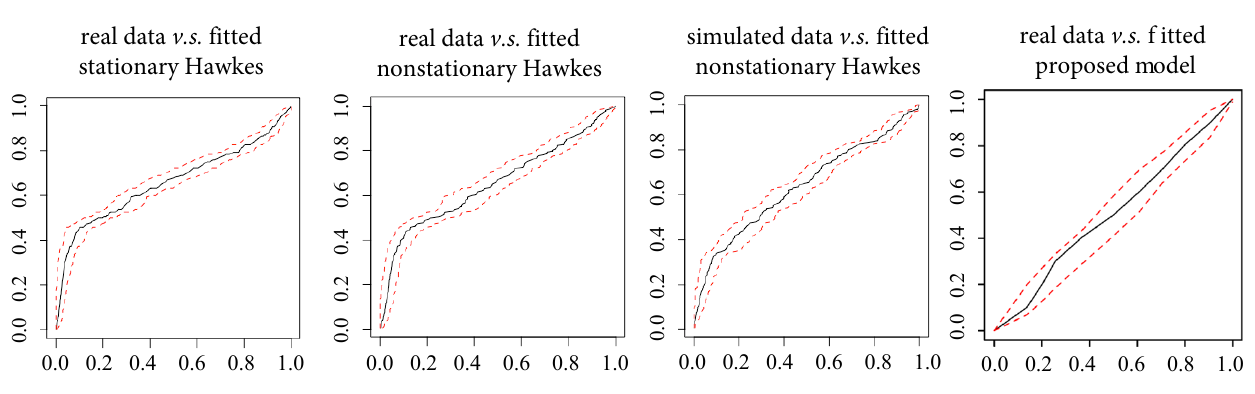}}
	\caption{Goodness-of-fit plots for Trump's Twitter data in January 2017. The plots from left to right show the empirical gap time distributions of the real data against those from the fitted stationary and nonstationary bivariate Hawkes processes, simulated realization from our proposed model against the fitted nonstationary Hawkes process, and the real data against the fitted proposed model.}
\label{trump}
\end{figure}
 
We fitted both stationary and nonstationary bivariate Hawkes process models to the two motivating datasets. The computational details are included in Section B.4 of the supplementary material. The first two plots in Figure~\ref{trump} show the goodness-of-fit plots for the models fitted to, for example, Trump's tweeting times in January 2017, the first month of his presidency. These plots were constructed by comparing the empirical distribution of gap time between two consecutive events against its theoretical counterpart from a fitted model. Based on these plots, we can see that the fitted Hawkes process models underestimated the gap time distribution, especially at small distances, leading to underestimated clustering strengths than in the data. 

These lacks of fit were likely caused by the potential inadequacy of Hawkes processes in modeling data with long intervals of inactivity between bursts of events, as suggested by \citet{Wu2019}. This point can be better appreciated by considering model \eqref{hawkes} in the stationary case. Intuitively, the intensity functions during any long interval of inactivity should be zero or at least extremely small. However, model \eqref{hawkes} suggests that the intensity functions at any time are bounded below by the constant background intensities which are larger than zero, assuming that the transfer functions are nonnegative. As a result, model \eqref{hawkes} will overestimate the true intensities in the long intervals of inactivity. Conversely, it will underestimate the activity rates outside such intervals. The problem can't be eliminated by simply using heterogeneous but nevertheless deterministic background intensities, because the long intervals of no activity are randomly scattered. 

A better modeling approach is to recognize the presence of potentially long intervals of no activity in the data and model them directly. We will develop one such model in the next section. The fourth plot in Figure~\ref{trump} is the goodness-of-fit plot based on our proposed model. No lack of fit is seen, which demonstrates the benefit of using our proposed model. In addition, we simulated a realization from our fitted model and then fitted a nonstationary bivariate Hawkes process model; the third plot in Figure~\ref{trump} is the resulting goodness-of-fit plot. It is interesting to see a similar lack-of-fit pattern to what was observed from modeling the real data, which supports our conclusion for the cause of poor fits from the Hawkes processes.

\section{Model formulation}

In our motivating application, there are two types of events (i.e., original posts and reposts). Hence, we focus on a bivariate point process model in our exposition. The proposed bivariate process can be easily reduced to a univariate process (see Section \ref{sec::dis}). To better illustrate the formulation, we describe our model specifications in the context of the social media data introduced in Section 1. 

Consider the observation window $[0,T]$. The observed event locations can be written as $\{T_1, T_2,\ldots,T_N\}$, where $0\le T_1<T_2<\ldots<T_N\le T$ and $N$ is a random variable taking nonnegative integer values. We assume that events arrive in \textit{non-overlapping} episodes (see Figure~\ref{illustrate}) and the first event of an episode is referred to as a \textit{parent} while the remaining events referred to as \textit{offsprings}. Write $[N]=\{1,\ldots,N\}$. Define a latent indicator variable $Y_l$, and let $Y_l=1$ ($Y_l=0$) if the $l$-th event is a parent (offspring), $l\in[N]$. For each event $l$, define a binary label $X_l$ such that $X_l=1$ ($X_l=0$) if it is an original post (repost), $l\in[N]$. 
{Note that event locations $\{T_l\}_{l\in[N]}$ and post/repost labels $\{X_l\}_{l\in[N]}$ are observed.
However, the parent/offspring labels $\{Y_l\}_{l\in[N]}$ are unobserved, as we do not know whether or not if an event is the first event of an episode. As such, the latent parent/offspring labels are treated as missing data in our model, analogously to the latent cluster label in mixture models.}

We assume that each episode contains alternating \textit{original post} and \textit{repost} segments (see Figure~\ref{illustrate}). An episode starts with an original post or a repost with probabilities $\alpha$ or $1-\alpha$. The number of segments in an episode is assumed to be $1+\text{Pois}(\gamma)$, and the numbers of events in an original post and a repost segment are assumed to be $1+\text{Pois}(\mu_1)$ and $1+\text{Pois}(\mu_0)$, respectively, where $\gamma, \mu_1, \mu_0>0$. Poisson distributions are used in our analysis but can be replaced by other distributions (e.g., geometric) generating nonnegative integers.
{Note that our model assumption allows an episode to contain only one event, which occurs when there is only one segment (with probability $e^{-\gamma}$) in an episode and this segment only contains one event (with probability $e^{-\mu_1}$ or $e^{-\mu_0}$), which is also the parent event.}

Let $D_l=T_l-T_{l-1}$, $l\in[N]$, be the gap times between adjacent events, where $T_0=0$. 
{Let $f_{l1}(d)$, $f^1_{l0}(d)$ and $f^0_{l0}(d)$ be the probability density functions of $D_l$ given $\{Y_l=1\}$, $\{Y_l=0, X_l=1\}$ and $\{Y_l=0,X_l=0\}$, respectively, and assume that $f_{l1}(d)$, $f^1_{l0}(d)$ and $f^0_{l0}(d)$ are parametric functions depending on some unknown parameters.} 
We assume that the probability of starting a new episode is time varying, and let
\begin{equation}
f_{l1}(d)=\lambda(t_{l-1}+d;\bbeta)\exp\left[-\int_{t_{l-1}}^{t_{l-1}+d} \lambda(t;\bbeta)\dd t\right],
\label{pargap}
\end{equation}
where $\lambda(t;\bbeta)$ is a parametric hazard function.
To avoid an overly complex model, we assume that the offspring gap times distributions $f^1_{l0}(d)$ and $f^0_{l0}(d)$ {(defined earlier in this paragraph)} are not functions of $t_l$; see Section~\ref{sec::dis} for more discussions. We assume that
\begin{equation}\label{expgap}
f_{l0}^1(d)=\rho_1\exp(-\rho_1d)\hbox{ and }  f_{l0}^0(d)=\rho_0\exp(-\rho_0d),
\end{equation}
where $\rho_1,\rho_0>0$ are unknown parameters.
Other distributions such as the Weibull distribution can also be considered.
One important characteristic of a user's content generating behavior is its strong daily cyclic pattern \citep{Guo2009}. To capture this characteristic, for example, we may model $\lambda(t;\bbeta)$ as a piece-wise polynomial function
\begin{equation}
\lambda(t;\bbeta)=\exp\left\{\sum_{i=1}^q\beta_iB_i(t-\floor{t})\right\},
\label{nonpara}
\end{equation}
where $B_1(\cdot),\ldots,B_q(\cdot)$ are cyclic B-spline basis functions defined on $[0,1]$ and $\bbeta=(\beta_1,\ldots,\beta_q)^\top$. 
{We remark that, while many processes, such as the Hawkes process in \eqref{hawkes}, are constructed through intensity functions, our model is formulated in terms of gap times. Specifically, given event labels $X_l$'s and $Y_l$'s, the gap times are assumed to independent. The dependence among event times is then introduced by integrating out the latent labels (i.e., $Y_l$'s) in the joint density function. An attractive feature of our model, formulated in terms of gap times, is that it enables us to characterize how a user may transition between active and inactive states and how bivariate events are generated during an active state.}

In the proposed model, the expected number of offsprings in an episode can be calculated as
\begin{equation}
\label{prop1}
\frac{1}{2}(2+\mu_1+\mu_0)(\gamma+1)+c(\gamma,\alpha)(\mu_1-\mu_0),
\end{equation}
where $c(\gamma,\alpha)=e^{-\gamma}(\alpha-1/2)\sum_{k=0}^{\infty}\gamma^{2k}/(2k)!$. Denote the expected gap times for offspring original post and repost as $e_1$ and $e_0$, respectively. The expected length of an episode is 
\begin{equation}
\label{prop2}
\frac{1}{2}\left[e_1(1+\mu_1)+e_0(1+\mu_0)\right](\gamma+1)+c(\gamma,\alpha)\left[e_1(1+\mu_1)-e_0(1+\mu_0)\right].
\end{equation}
The proof of \eqref{prop1} and \eqref{prop2} is given in Section A.1. of the supplementary material.

\section{Estimation}
In the proposed model, event locations $\t=\{t_1,\ldots,t_n\}$ and original post/repost labels $\x=\{x_1,\ldots,x_n\}$ are observed. 
However, parent/offspring labels $\y=\{y_1,\ldots,y_n\}$ are not observed, as we do not know whether or not if an event is the first event of an episode. Write the total number of episodes as $K$ (i.e., $K=\sum_{l=1}^ny_l$), the number of segments in the $k$-th episode as $n_k$ and the number of events in the $j$-th segment of the $k$-th episode as $l_{k_j}$, $j\in[n_k]$, $k\in[K]$. Define an indicator $z_{k_j}$ such that $z_{k_j}=1$ if the $j$-th segment in the $k$-th episode is an original post segment and $z_{k_j}=0$ otherwise, $j\in[n_k]$, $k\in[K]$.

Write $\btheta=\{\alpha,\gamma,\mu_1,\mu_0,\rho_1,\rho_0,\bbeta\}$. Assume that the first event is a parent event and all events of the last episode are contained in $[0,T]$. 
The observed-data likelihood function, by treating $\y$ as missing data, can be written as
\begin{equation}
L(\btheta;\t,\x)=\sum_{\y\in\mathcal{Y}}f(\t,\x,\y|\btheta),
\label{like}
\end{equation}
where $\mathcal{Y}$ is the set of all binary vectors of length $n$ with $y_1=1$. The joint density of $\t$, $\x$ and $\y$ given $\btheta$ is written as
\begin{eqnarray}
\label{full}
f(\t,\x,\y|\btheta)&=&\prod_{l=1}^nf_{l}(d_l)\times P(D_{n+1}>T-t_n)\times\prod_{l=1}^n\alpha^{I(y_l=1,x_l=1)}(1-\alpha)^{I(y_l=1,x_l=0)}\\\nonumber
&&\times\prod_{k=1}^K\frac{\gamma^{n_k-1}e^{-\gamma}}{(n_k-1)!}\times\prod_{k=1}^K\prod_{j=1}^{n_k}\frac{(\mu_1^{l_{k_j}-1}e^{-\mu_1})^{I(z_{k_j}=1)}(\mu_0^{l_{k_j}-1}e^{-\mu_0})^{I(z_{k_j}=0)}}{(l_{k_j}-1)!},
\end{eqnarray} 
where $f_{l}(d_l)=f_{l1}(d_l)^{I(y_l=1)}f^1_{l0}(d_l)^{I(y_l=0,x_l=1)}f^0_{l0}(d_l)^{I(y_l=0,x_l=0)}$, $d_l=t_l-t_{l-1}$, $l\in[n]$, and $D_{n+1}$ is the gap time between $t_n$ and the next parent event. With straightforward algebra, we have
$
P(D_{n+1}>T-t_n)=\exp\left[-\int_{t_n}^{T} \lambda(t;\bbeta)\dd t\right].
$
{To ease the notation, when summing over all possible $\y$'s in $\mathcal{Y}$, we write it as $\sum_{\y}$ without emphasizing that $\y\in\mathcal{Y}$.}

To estimate $\btheta$, directly maximizing the likelihood function in (\ref{like}) is computationally impractical since the number of elements in $\mathcal{Y}$ grows exponentially with $n$.
An alternative approach is to employ an EM algorithm that treats $Y_1,\ldots, Y_n$ as missing data.
However, the E-step in the EM procedure requires calculating $P_{\btheta}(Y_l=h|\t,\x)$, which is not tractable.
To overcome the computational difficulty, we consider a composite likelihood approach in the next section.

\subsection{Composite likelihood}
\label{sec::clike}
The composite likelihood approach makes statistical estimation and inference through an inference function derived by multiplying a collection of component likelihoods \citep{Lindsay1988}. 
Write the length of a sub-window as $s\in\mathbb{R}$. We divide $[0,T]$ into $M$ nonoverlapping sub-windows, i.e., $[0,T]=\bigcup_{m=1}^M[ms-s,ms)$. Define $\t_m=\{t_i:t_i\in[ms-s,ms), i\in[n]\}$, $m\in[M]$. We use binary vectors $\y_m$ and $\x_m$ to indicate the parent/offspring events and original posts/reposts in $\t_m$, respectively.

In each sub-window, we assume that the first event is a parent event and all the events in the last episode are contained in the sub-window. Neither assumption is restrictive from a practical point of view. A typical user is inactive during some fixed time interval at night and such an interval can be identified by examining the posting times. By setting day as the sub-window, a post made right before or after that interval is therefore the last event from the previous episode or a parent event for the next episode. These assumptions allow for a fast calculation of the composite likelihood function, and the approximation bias is negligible when the number of windows $M$ is not too large; see Theorem~\ref{thm3} for details. Under these assumptions, we can write the composite likelihood function as $L_s^c(\btheta;\t,\x)=\prod_{m=1}^M f(\t_m,\x_m|\btheta)$,
where $f(\t_m,\x_m|\btheta)=\sum_{\y_m} f(\t_m,\x_m,\y_m|\btheta)$ and $f(\t_m,\x_m,\y_m|\btheta)$ is defined as in (\ref{full}). 
Hence, the log composite likelihood function can be written as
\begin{equation}
\ell_s^c(\btheta;\t,\x)=\frac{1}{T}\sum_{m=1}^M\log\left[\sum_{\y_m} f(\t_m,\x_m,\y_m|\btheta)\right].
\label{com}
\end{equation}
{Calculating $\sum_{\y_m} f(\t_m,\x_m,\y_m|\btheta)$ requires summing over $2^{|\y_m|-1}$ terms, $m\in[M]$, where $|\cdot|$ denotes the number of elements in a vector or a set.
When $\sup_m|\y_m|\ll |\y|$, the calculation can be performed much more efficiently. 
Moreover, the computation cost $O(\sum_{m=1}^M 2^{|\y_m|})$ only increases approximately linearly with the observation window length $T=Ms$.}

\subsection{CLEM algorithm}
To estimate $\btheta$, a straightforward approach is to directly maximize \eqref{com} using numerical methods. This approach is not desirable as it suffers from low computational efficiency and stability. First, in the numerical methods, each step involves maximizing \eqref{com} with respect to $\btheta=(\alpha,\gamma,\mu_1,\mu_0,\rho_1,\rho_0,\bbeta)$ {under constraints such as} $0<\alpha<1$ and $\gamma,\mu_1,\mu_0>0$, which is computationally costly. Second, objective functions such as that in \eqref{com} often has a flat surface \citep{Veen2008}. As such, both the computation time and the convergence can be sensitive to the starting values.

We propose an efficient and stable CLEM algorithm that requires calculating only $\y_m|\t_m,\x_m$ (as opposed to $\y|\t,\x$) for a given $\btheta$, $m\in[M]$.
To that end, define
$$
Q(\btheta|\btheta_{p-1})=\frac{1}{T}\sum_{m=1}^ME_{\Y_m|\t_m,\x_m,\btheta_{p-1}}\log \left[f(\t_m,\x_m,\Y_m|\btheta)\right],
$$
where $\btheta_{p-1}$ is the  update after completing the $(p-1)$-th iteration.
The CLEM algorithm iterates between the following two steps until convergence.
\begin{itemize}
\item E-step: Given the previous update $\btheta_{p-1}$, obtain $Q(\btheta|\btheta_{p-1})$.
\item M-step: Maximize $Q(\btheta|\btheta_{p-1})$ with respect to $\btheta$ to produce $\btheta_p$.
\end{itemize}

{In the E-step, {we will need to evaluate} the conditional distribution $P_{\btheta}(Y_l=h|\t_m,\x_m,\btheta)$ {which} is calculated as
\begin{eqnarray}\label{eq:condP1}
P_{\btheta}(Y_l=h|\t_m,\x_m,\btheta)={{\sum_{\y_m|y_l=h} f(\t_m,\x_m,\y_m|\btheta)\over \sum_{\y_m} f(\t_m,\x_m,\y_m|\btheta)}},\quad h=0,1.
\end{eqnarray}
Due to the composite EM formulation, this is much easier to calculate because the summations in (\ref{eq:condP1}) are over $\y_m$ rather than $\y$.
When $\sup_m|\y_m|\ll |\y|$, the calculation can be performed much more efficiently. 
It is seen that the computation cost only increases approximately linearly with the observation window length $T=Ms$.}
{Moreover, if we can identify several events as parent events a priori, the computational complexity can be further reduced. This ``speedup" procedure is detailed in Section B.2.}
In the M-step, all parameters except for $\bbeta$ have closed-form updates. Hence, the M-step can be achieved efficiently. More computational details on the CLEM algorithm can be found in the supplementary material. While we do not observe notable differences in the parameters estimated from the CLEM algorithm and directly applying numerical methods (in the cases that they do converge), numerical methods on average take more than 15 times longer to reach convergence in our simulation studies. 

In Section \ref{sec:theory}, we show that the CLEM algorithm, although working only with $\y_m|\t_m,\x_m$ (as opposed to $\y|\t,\x$), still enjoys the desirable ascent property, which guarantees that the log composite likelihood is non-decreasing at each CLEM iteration. Since $Q(\btheta|\btheta')$ is continuous in both $\btheta$ and $\btheta'$, the convergence of $\btheta_p$ to a stationary point as $p\rightarrow\infty$ is guaranteed by Theorem 2 in \cite{Wu1983}. Whether it converges to a global or local maximum depends on the initial value. Common techniques such as running the algorithm from multiple starting points can help locate the global maximum.

\subsection{Goodness of fit}
\label{sec::good}

Having estimated the parameters in the proposed model, it is important to assess whether or not the estimated model fits the point patterns observed in the data. Residual analysis-based assessment \citep{Baddeley2005} cannot be applied to our setting since our model is formulated using gap times and its intensity function is very difficult to derive. Alternatively, we evaluate the goodness of fit by checking whether the fitted model can adequately capture the inhomogeneity in the gap times calculated from the observed data.
We propose a goodness-of-fit procedure that compares the empirical gap time distribution to that calculated from realizations simulated from the fitted model. Details of the procedure are included in Section B.3 of the supplementary material.

\section{Theoretical properties}
\label{sec:theory}
In this section, we show the convergence guarantee of the CLEM algorithm, the consistency and asymptotic normality of the estimator. We also discuss the estimation of the variance-covariance matrix in practice. 
We use $\btheta_0$ to denote the true parameter vector, $\bm{\Theta}$ to denote the parameter space for $\btheta$, and assume that $\bm{\Theta}$ is compact.

First, we show that the above CLEM algorithm enjoys the desirable ascent property, which guarantees that the log composite likelihood is non-decreasing at each CLEM iteration. The proof is given in Section A.2. of the supplementary material.
\begin{theorem}
The composite log-likelihood $\ell_s^c(\btheta;\t,\x)$ and the CLEM sequence $\btheta_p$ satisfy
$$
\ell_s^c(\btheta_p;\t,\x)\ge\ell_s^c(\btheta_{p-1};\t,\x),
$$
where the equality holds if and only if $Q(\btheta_{p}|\btheta_{p-1})=Q(\btheta_{p-1}|\btheta_{p-1})$, $p=1,2,\ldots$.
\label{thm1}
\end{theorem}

In the ensuing theoretical development, we assume that $\hat\btheta_{M,s}$ is the maximizer of the composite likelihood estimator.
Consider the log composite likelihood function in (\ref{com}). 
Its composite score function can be written as
\[
U_{M,s}(\btheta)={\frac{1}{T}\sum_{m=1}^MU_{m,s}(\btheta)},\quad\text{where}\quad U_{m,s}(\btheta)=\frac{f^{(1)}(\t_m,\x_m|\btheta)}{f(\t_m,\x_m|\btheta)},
\]
and $f^{(1)}(\t_m,\x_m|\btheta)$ is the first-order derivative with respect to $\btheta$.
The maximum composite likelihood estimator $\hat\btheta_{M,s}$ in our proposed method is the solution to $U_{M,s}(\btheta)=0$.
Here, we write $U_{M,s}$ to signify that this score function is also a function of the sub-window length $s$.
In the next theorem, we establish consistency of $\hat\btheta_{M,s}$. 
\begin{theorem}
\label{thm3}
Assume that the following conditions are satisfied,
\begin{description}
\item{(2.1)} $E[U_{M,s}(\btheta)]=0$ only at $\btheta=\btheta_s^*$,
\item{(2.2)} There exists a nonnegative function $\kappa(\cdot)$ such that 
$$
\left|\frac{f^{(1)}(\t_m,\x_m|\btheta)}{f(\t_m,\x_m|\btheta)}\right|<\kappa(|\t_m|),\quad \left|\frac{f^{(2)}(\t_m,\x_m|\btheta)}{f(\t_m,\x_m|\btheta)}\right|<\kappa(|\t_m|),
$$
and $E[\kappa(|\t_m|)^2]<\infty$. Here, $|\t_m|$ is the number of events in the $m$-th sub-window. 
\end{description}
Then, we have $\hat\btheta_{M,s}$ converges in probability to $\btheta^*_s$ as $M\rightarrow\infty$. Moreover, if $E[U_{M,s}(\btheta)]\rightarrow0$ as $s\rightarrow\infty$ only at $\btheta=\btheta_0$, we have $\btheta^*_s\rightarrow\btheta_0$ as $s\rightarrow\infty$.
\end{theorem}
\noindent
The proof is given in Section A.3 of the supplementary material. In the theorem, we first show that $\hat\btheta_{M,s}$ converges to $\btheta^*_s$ as the number of sub-windows tends to infinity.
As the density $f(\t_m,\x_m|\btheta)$ is calculated by assuming that the first event in a sub-window is a parent event and all events of the last episode are contained in the sub-window are satisfied, it is only an approximation to the true density function.
Note both assumptions used in the approximation involve only the first and the last episodes in the sub-window. Thus, $\btheta^*_s$ converges to $\btheta_0$ as the length of the sub-window increases. This rate of convergence is investigated in Theorem 3.

In the next lemma, we show that the model specifications considered in our work satisfy Conditions (2.1) and (2.2). The proof is given in Section A.4 of the supplementary material.

\begin{lemma}
\label{lem1}
In our proposed model, when the parent hazard function $\lambda(t,\bbeta)$ satisfies
$$
\max_{j}\left|\frac{\partial\lambda(t,\bbeta)/\partial\beta_j}{\lambda(t,\bbeta)}\right|<\infty,
$$
we have that Conditions (2.1) and (2.2) in Theorem~\ref{thm3} are satisfied. Consequently, the consistency result in Theorem~\ref{thm3} applies to our composite likelihood estimator.
\end{lemma}

{As we discussed earlier, the density $f(\t_m,\x_m|\btheta)$ is an approximation to the true density, denoted as $f_0(\t_m,\x_m|\btheta)$, whose exact form is given in \eqref{eqn:f0} of the supplementary material. The difference between $f(\t_m,\x_m|\btheta)$ and $f_0(\t_m,\x_m|\btheta)$ is due to that $f(\t_m,\x_m|\btheta)$ is calculated by assuming the first event in a sub-window is a parent event and all events of the last episode are contained in the sub-window. As discussed in Section \ref{sec::clike}, the use of $f(\t_m,\x_m|\btheta)$ allows for a considerable more efficient calculation in the M-step of the CLEM algorithm, and the approximation bias is negligible as long as the number of windows $M$ is not too large; see Theorem~\ref{thm3} for details.}
Define $\mathcal{H}_0(\btheta)=-\mathbb{E}\left[\frac{\partial}{\partial \btheta} U_{0}(\btheta)\right]$, where 
$U_{0}(\btheta)=\frac{1}{T}\sum_{m=1}^M\frac{f_0^{(1)}(\t_m,\x_m|\btheta)}{f_0(\t_m,\x_m|\btheta)}$.
Next, we show the asymptotic normality of $\hat\btheta_{M,s}$ as $T\rightarrow\infty$.

\begin{theorem}
\label{thm3}
Assume conditions in Theorem 2 hold and {$M=O(T^{2/5})$}. Define
$$
\mathcal{H}(\btheta_0)=-\left. \left\{\mathbb{E}\left[\frac{\partial}{\partial \btheta}U_{M,s}(\btheta)\right]\right\}\right|_{\btheta=\btheta_0},
$$
$$
\mathcal{I}_M^{-1}=\mathcal{H}^{-1}(\btheta_0)\left\{{\frac{1}{T}\sum_{i=1}^M\sum_{j=1}^M \mathbb{E}\left[(U_{i,s}(\btheta_0)U^T_{j,s}(\btheta_0)\right]}\right\}\mathcal{H}^{-1}(\btheta_0)^{T}.
$$
If $\mathcal{H}(\btheta)$ and $\mathcal{H}_0(\btheta)$ are positive definite for $\btheta\in\mathcal{B}_{r_0}(\btheta_0)$ and some constant $r_0>0$, {where $\mathcal{B}_{r_0}(\btheta_0)$ denotes the Frobenius-norm ball around $\btheta_0$ with radius $r_0>0$,} then {$\sqrt{T}(\hat\btheta_{M,s}-\btheta_0)$ converges in distribution to $N(\bm{0},\mathcal{I}_M^{-1})$}.
\end{theorem}
The proof is given in Section A.5 of the supplementary material. 
{
The condition on $\mathcal{H}(\btheta)$ and $\mathcal{H}_0(\btheta)$ requires the composite likelihoods to be strongly convex in a small region around $\btheta_0$. Theorem~\ref{thm3} ensures that the inference for the true unknown parameter $\btheta_0$ is valid as long as $M=O(T^{2/5})$. 
As the computational cost decreases with $M$ (see discussion in Section \ref{sec::clike}), we suggest choosing a large $M$ such as $M=cT^{2/5}$ for some constant $c>0$ in practical implementations. In Section \ref{sec::sim}, we demonstrate that the estimation accuracy is not sensitive to the choice of $M$.}

In practice, the variance-covariance matrix $\mathcal{I}^{-1}_M$ needs to be estimated.
If we assume that the $U_{m,s}(\btheta)$'s are independent, we can estimate $\mathcal{H}$ and $\mathcal{I}_M^{-1}$ using
$$
\mathcal{\hat I}_0=-\left. \left[\frac{1}{T}\sum_{i=1}^M\frac{\partial}{\partial \btheta}U_{m,s}(\btheta)\right]\right|_{\btheta=\hat\btheta_{M,s}},
$$
$$
\mathcal{\hat I}_M^{-1}=(\mathcal{\hat I}^{-1}_0)\left\{\frac{1}{T}\sum_{m=1}^M \left[U_{m,s}(\hat\btheta_{M,s})U^T_{m,s}(\hat\btheta_{M,s})\right]\right\}(\mathcal{\hat I}^{-1}_0)^{T},
$$
where $U_{m,s}(\hat\btheta_{M,s})$ and $\frac{\partial}{\partial \btheta}U_{m,s}(\btheta)|_{\btheta=\hat\btheta_{M,s}}$, $m\in[M$], can be estimated in the last step of the CLEM procedure.
{Without the independence assumption on the $U_{m,s}(\btheta)$'s, we need to adopt a simulation approach to estimate the following term that appears in the covariance formula
$\mathcal{J}_M=\frac{1}{T}\sum_{i=1}^M\sum_{j=1}^M \mathbb{E}\left[(U_{i,s}(\btheta_0)U^T_{j,s}(\btheta_0)\right]$.
Specifically, for the $h$th realization simulated from the estimated model with parameter $\hat\btheta_{M,s}$, we can estimate $\{\hat U^{(h)}_{i,s}\}_{i\in[M]}$. Given $q$ realizations, we can then estimate $\mathcal{J}_M$ with
$$
\mathcal{\hat J}_M=\frac{1}{qT}\sum_{h=1}^q\sum_{i=1}^M\sum_{j=1}^M\hat U^{(h)}_{i,s}\hat U^{(h)}_{j,s}{}^\top.
$$
We further demonstrate the efficacy of this estimation approach in Section \ref{sec::sim}.}

\section{Simulation study}\label{sec::sim}
We simulate point processes from the proposed model with gap time distributions given in (\ref{pargap}) and (\ref{expgap}) with
\begin{equation}\label{simgap}
\lambda(t;\bbeta)=\exp\left[\beta_{01}+\beta_{11}\cos(2\pi t)+\beta_{12}\sin(2\pi t)\right]
\end{equation}
and $\bbeta=(\beta_{01},\beta_{11},\beta_{12})^\top$.
To simulate data from the model, we use the thinning technique proposed in Lewis and Shedler (1979).
We set the observation window length $T=100$, $\alpha=0.6$, $\gamma=0.5$ or $1$, $\mu_1=0.5$ or $1$, $\mu_0=0.5$ or $1$, $(\rho_1,\rho_0)=(10,15)$ or $(20,30)$ and $\bbeta^\top=(-2,-2,2)$ or $(-3,-3,3)$.
With each parameter configuration, we simulate 100 event trajectories. 
For estimation, we use sub-window length $s=5$ (or $M=20$).
Furthermore, to model $\lambda(t,\bbeta)$, we consider both the true model in \eqref{simgap} and the nonparametric cyclic B-spline model in (\ref{nonpara}). For the latter, we use the 9 equally spaced knots in $[0,1]$. 

\begin{table}[!htb]
\caption{\label{t1}{Parameter estimation using \eqref{simgap} for $\lambda(t,\bbeta)$, with processes simulated under $\alpha=0.6$ and different $(\gamma,\mu_1,\mu_0,\rho_1,\rho_0,\bbeta)$. Under each setting, the simulated standard errors are given in the second row and the estimated standard errors are given in the third row.}}
\setlength{\tabcolsep}{1.75pt}
\centering
{\begin{tabular}{c|ccccccccc}
\hline$(\gamma,\mu_1,\mu_0,\rho_1,\rho_0)$\\$(\beta_{01},\beta_{11},\beta_{12})$ &$\alpha$ & $\gamma$ & $\mu_1$ & $\mu_0$ & $\rho_1$ & $\rho_0$ & $\beta_{01}$ & $\beta_{11}$ & $\beta_{12}$ \\ \hline
(0.5,0.5,0.5,10,15)    & 0.597 &0.510  &0.497 &0.501 &10.690 &15.677 &-1.963 &-1.962 &1.969       \\
 (-2,-2,2)  &(0.008) &(0.011) &(0.011) &(0.014) &(0.144) &(0.341)   &(0.045)  &(0.036) &(0.046)      \\ 
 & (0.007) &(0.012) &(0.012) &(0.013) &(0.193) &(0.302) &(0.049) &(0.039) &(0.053)\\ \hline
 (0.5,0.5,0.5,10,15) & 0.599 &0.505 &0.498 &0.500 &10.407 &15.822 &-2.933 &-2.890 &2.910        \\
 (-3,-3,3) & (0.007) &(0.010) &(0.011) &(0.010) &(0.183) &(0.270)  &(0.085) &(0.064) &(0.074)\\ 
 &(0.007) &(0.012) &(0.012) &(0.013) &(0.182) &(0.280) &(0.073) &(0.058) &(0.071)\\ \hline
 (1.0,0.5,0.5,10,15)   & 0.600 &0.915 &0.493 &0.479 &10.806 &16.213 &-2.015 &-2.006 &2.020     \\
 (-2,-2,2) & (0.006) &(0.015) &(0.011) &(0.009) &(0.156) &(0.271)  &(0.058) &(0.054) &(0.050)\\ 
 &(0.007) &(0.016) &(0.011) &(0.011) &(0.159) &(0.253) &(0.052) &(0.044) &(0.056)\\ \hline
  (0.5,1.0,1.0,10,15)   & 0.605 &0.488 &0.953 &0.947 &10.880 &15.768 &-2.058 &-1.970 &2.026   \\
 (-2,-2,2) & (0.008) &(0.007) &(0.014) &(0.016) &(0.144) &(0.251)  &(0.055) &(0.046) &(0.056)\\ 
 &(0.007) &(0.012) &(0.017) &(0.018) &(0.159) &(0.244) &(0.053) &(0.044) &(0.059)\\ \hline
  (0.5,0.5,0.5,20,30)   & 0.594 &0.505 &0.509 &0.516 &20.805 &31.698 &-2.159 &-2.112 &2.127        \\
 (-2,-2,2) & (0.007) &(0.013) &(0.012) &(0.014) &(0.408) &(0.741)  &(0.050) &(0.044) &(0.045)\\ 
 &(0.007) &(0.012) &(0.013) &(0.013) &(0.439) &(0.699) &(0.047) &(0.036) &(0.051)\\ \hline
\end{tabular}}
\end{table}

\begin{table}[!t]
\caption{\label{t2}{Parameter estimation using \eqref{nonpara} for $\lambda(t,\bbeta)$, with processes simulated under $\alpha=0.6$ and different $(\gamma,\mu_1,\mu_0,\rho_1,\rho_0,\bbeta)$. Under each setting, the simulated standard errors are given in the second row and the estimated standard errors are given in the third row.}}

\setlength{\tabcolsep}{1.75pt}
\centering
{\begin{tabular}{c|cccccc}
\hline$(\gamma,\mu_1,\mu_0,\rho_1,\rho_0)$\\$(\beta_{01},\beta_{11},\beta_{12})$ &$\alpha$ & $\gamma$ & $\mu_1$ & $\mu_0$ & $\rho_1$ & $\rho_0$ \\ \hline
{(0.5,0.5,0.5,10,15)} & 0.594 &0.475  &0.514 &0.478 &11.176 &16.340 \\
(-2,-2,2) &(0.007) &(0.011) &(0.014) &(0.014) &(0.268) &(0.364)    \\ 
 & (0.008) &(0.014) &(0.014) &(0.014) &(0.269) &(0.396)\\ \hline
{(0.5,0.5,0.5,10,15)} & 0.600 &0.513  &0.490 &0.489 &11.013 &15.583 \\
(-3,-3,3) &(0.007) &(0.013) &(0.013) &(0.011) &(0.163) &(0.253)    \\ 
  & (0.007) &(0.013) &(0.012) &(0.013) &(0.198) &(0.291)\\ \hline
(1.0,0.5,0.5,10,15) & 0.595 &0.922  &0.478 &0.499 &11.091 &15.527 \\
(-2,-2,2) &(0.007) &(0.013) &(0.010) &(0.010) &(0.197) &(0.248)    \\ 
 & (0.009) &(0.022) &(0.012) &(0.012) &(0.216) &(0.273)\\ \hline
(0.5,1.0,1.0,10,15) & 0.615 &0.436  &0.921 &0.957 &10.895 &15.982 \\
(-2,-2,2) &(0.008) &(0.010) &(0.018) &(0.019) &(0.153) &(0.254)    \\ 
 & (0.008) &(0.015) &(0.021) &(0.020) &(0.219) &(0.351)\\ \hline 
 (0.5,0.5,0.5,20,30)  & 0.588 &0.533 &0.517 &0.510 &20.094 &31.279\\
 (-2,-2,2) & (0.007) &(0.012) &(0.013) &(0.011) &(0.354) &(0.701)  \\ 
 &(0.008) &(0.014) &(0.013) &(0.013) &(0.493) &(0.645)\\ \hline
\end{tabular}}
\end{table}

\begin{figure}[!t]
\centering
\includegraphics[trim=-1cm 0 0cm 0, scale=0.65]{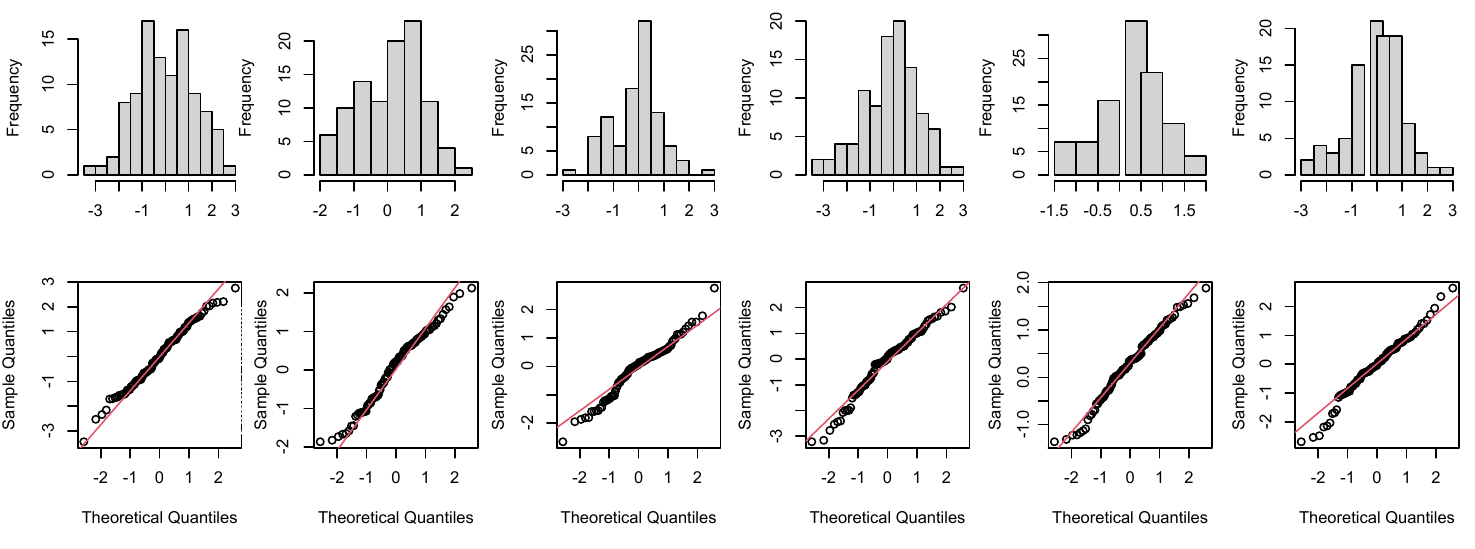}
\caption{{Histograms and QQ plots of estimated parameters $(\alpha,\gamma,\mu_1,\mu_0,\rho_1,\rho_0)$ standardized by the estimated asymptotic variance.}}
\label{fig3.3}
\end{figure}

Table~\ref{t1} and Table~\ref{t2} show the parameter and standard error estimates when $\lambda(t,\bbeta)$ is specified as in \eqref{simgap} and \eqref{nonpara}, respectively.
In both tables, the estimated parameters are close to the true values. 
With all other parameters fixed, models with $\bbeta^\top=(-3,-3,3)$ generate more episodes and segments compared to models with $\bbeta^\top=(-2,-2,2)$. Therefore, parameters $\gamma$, $\mu_1$, $\mu_0$, $\rho_1$ and $\rho_0$ are estimated better when $\bbeta^\top=(-3,-3,3)$.
This can be observed by comparing the standard errors in the first and second settings in Table~\ref{t1} (or Table~\ref{t2}).
With all other parameters fixed, greater $\gamma$ leads to longer episodes with more offspring and, therefore, better estimations of $\mu_1$, $\mu_0$, $\rho_1$ and $\rho_0$.
We can observe this by comparing the standard errors in the first and the third settings in Table~\ref{t1} (or Table~\ref{t2}). 
Comparing the estimates of $\gamma,\mu_1,\mu_0,\rho_1,\rho_0$ in Table~\ref{t1} and Table~\ref{t2}, we can see that estimating $\lambda(t,\bbeta)$ using B-splines gives satisfactory performance even though the true underlying {hazard} function is exponential sinusoidal. 
{It is also seen from Tables \ref{t1} and \ref{t2} that the estimated asymptotic standard errors (i.e., standard errors estimated using the asymptotic formula) are close to the simulated standard errors (i.e., standard errors estimated using estimates from data replicates). In Figure \ref{fig3.3}, we plot the histograms and the QQ plots of estimated parameters $(\alpha,\gamma,\mu_1,\mu_0,\rho_1,\rho_0)$ standardized by the estimated asymptotic variance, under the setting in the first row of Table \ref{t1} over 100 data replicates. It is seen that the empirical distributions are in good agreement with the standard normal density.
We also performed Kolmogorov-Smirnov tests for all such standardized parameter estimates against the standard normal distribution, and the p-values are greater than 0.05 for all settings considered in Table \ref{t1}.
}

{Furthermore, we evaluate the sensitivity of the our method to the choice of $M$. Under the same setting as in Table \ref{t1} where
$(\gamma,\mu_1,\mu_0,\rho_1,\rho_0;\bbeta)=(0.5,0.5,0.5,10,15;-2,-2,2)$ and $T=100$, we consider setting the number of sub-windows $M=10,20,50,100$. The results over 100 data replicates are shown in Table \ref{tab.1} and it is seen that the estimation accuracy is not overly sensitive to the choices of $M$.
We can see some bias when $M=100$. However, when the number of sub-windows is reduced to $M=50$, we see no evidence of bias and the estimated parameters are close to the true values. When the number of sub-windows is further decreased, we see no noticeable difference in the results.
This finding is consistent with our theoretical results.}

\begin{table}[!t]
\caption{\label{tab.1}{Parameter estimation with data simulated under $\alpha=0.6$ and  $(\gamma,\mu_1,\mu_0,\rho_1,\rho_0,\bbeta)=(0.5,0.5,0.5,10,15;-2,-2,2)$ when $M=10,20,50,100$. The standard errors are given in parentheses.}}
\centering
{\setlength{\tabcolsep}{1.75pt}
\begin{tabular}{c|ccccccccc}
\hline &$\alpha$ & $\gamma$ & $\mu_1$ & $\mu_0$ & $\rho_1$ & $\rho_0$ & $\beta_{01}$ & $\beta_{11}$ & $\beta_{12}$ \\ \hline
$M=10$& 0.598 &0.508  &0.499 &0.505 &10.953 &15.460 &-2.051 &-2.002 &2.033       \\
&(0.007) &(0.011) &(0.012) &(0.012) &(0.206) &(0.315)   &(0.048)  &(0.041) &(0.048)      \\ \hline
$M=20$ & 0.597 &0.510  &0.497 &0.501 &10.690 &15.677 &-1.963 &-1.962 &1.969       \\
&(0.008) &(0.011) &(0.011) &(0.014) &(0.144) &(0.341)   &(0.045)  &(0.036) &(0.046)      \\  \hline
$M=50$& 0.596 &0.488  &0.500 &0.488 &10.924 &15.934 &-2.063 &-2.027 &2.051       \\
&(0.007) &(0.012) &(0.011) &(0.013) &(0.215) &(0.305)   &(0.047)  &(0.042) &(0.047)      \\ \hline 
$M=100$& 0.599 &0.480  &0.451 &0.484 &10.918 &16.214 &-1.801 &-1.872 &1.857       \\
&(0.008) &(0.011) &(0.012) &(0.012) &(0.198) &(0.329)   &(0.041)  &(0.039) &(0.040)      \\ \hline
\end{tabular}}
\end{table}

\section{Social media data analysis}
\label{sec:real}
In this section, we apply our proposed model to the two social media datasets. In the first application, we study Twitter data collected from Donald Trump from January 2013 to April 2018, and characterize changes in various aspects of his tweeting behavior, such as the tweeting rate, length of each tweeting episode and daily activity level, before, during and after the presidential campaign. In the second application, we apply our proposed method to a large-scale user data collected from Sina Weibo. Through investigating different aspects of user behaviors, we find interesting user subgroups. Furthermore, we discuss the effect of social ties on a user's posting behavior.

\subsection{Donald Trump twitter data} 
\label{sec::trump}
We study the Twitter data collected from Donald from Donald Trump, the 45th and current President of the United States. The data were collected from Donald Trump's personal twitter account @realDonaldTrump. An archive of all tweets published from this account can be downloaded at \url{http://www.trumptwitterarchive.com/}. We focus on the time period from January 2013 to April 2018. The average number of monthly tweets is 278 with a standard deviation of 154. 

\begin{figure}[!t]
\centering
\includegraphics[width=1\textwidth]{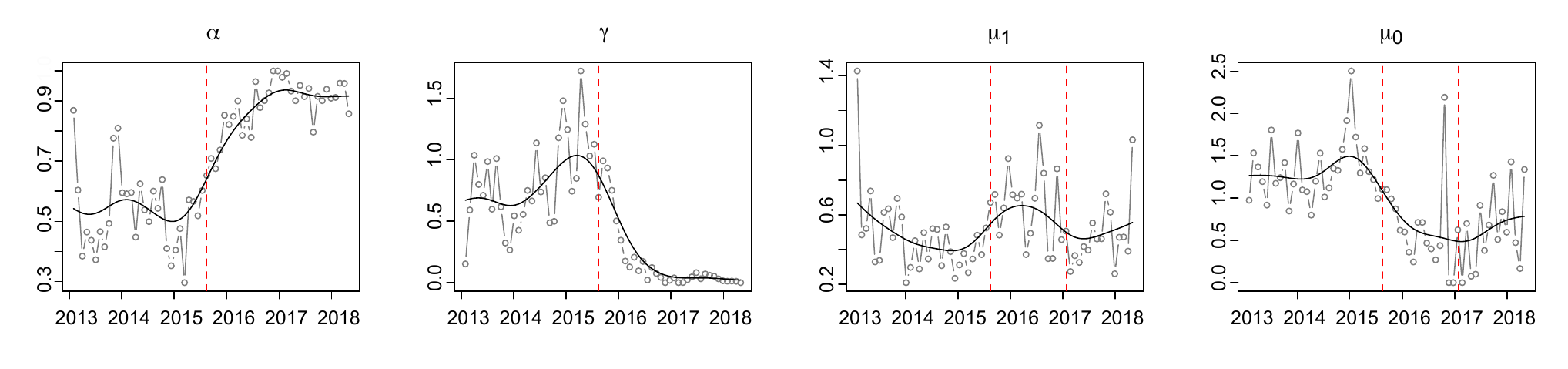}
\includegraphics[width=1\textwidth]{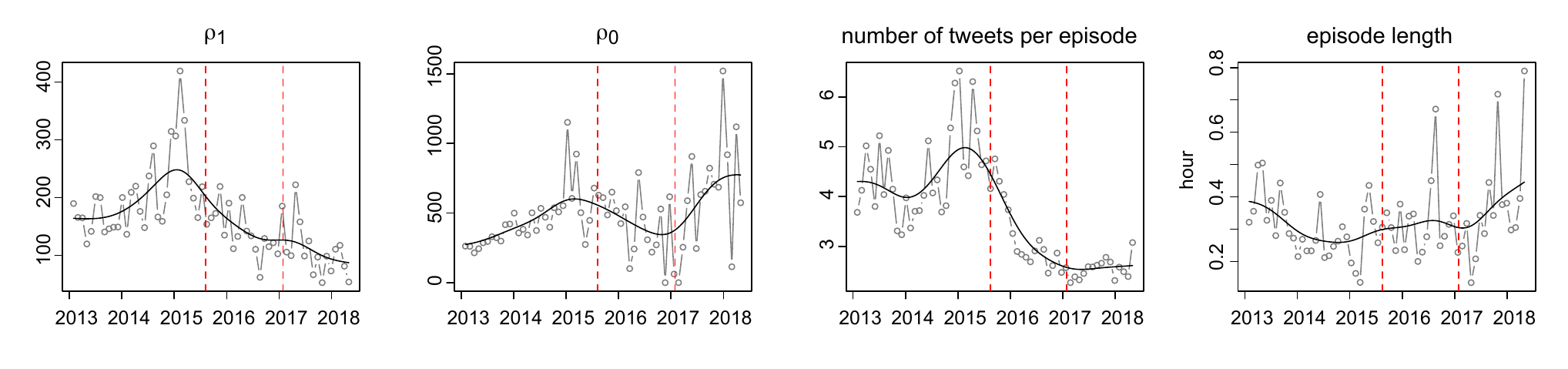}
\caption{Parameters estimated from Donald Trump's monthly Twitter data. The plotted points show estimated parameters for each month and the black solid lines show the Gaussian kernel smoothed curves. The two red dashed lines mark June 2015 (candidacy announcement) and January 2017 (assumed office), respectively.}
\label{trump1}
\end{figure}

We fit the proposed model to the tweets collected within each month in the study window. We model the offspring gap times using (\ref{expgap}), and the parent hazard function using (\ref{nonpara}) with 7 equally spaced knots in one day. We consider a sub-window with length $s=7$ days. The estimated parameters are shown in Figure~\ref{trump1}, in which two important months are marked. The first one is June 2015, the month in which Trump announced his candidacy for president; the second one is January 2017, the month in which he had the inauguration and assumed office. The interpretations of the plotted parameters are summarized as follows:
\begin{table}[!htb]
\begin{tabular}{ll}
$\alpha$ & the probability that an episode starts with an original post,                   \\
$\gamma$ & the average number of switches between segments in an episode, \\
$\mu_1+1$  & the average number of posts in an original post segment,                                \\
$\mu_0+1$  & the average number of posts in a repost segment,                                       \\
$\rho_1$ & the original post rate (rate parameter in the exponential distribution),                                                         \\
$\rho_0$ & the repost rate (rate parameter in the exponential distribution).                                                               
\end{tabular}
\end{table}

\begin{figure}[!t]
\centering
\includegraphics[width=0.55\textwidth]{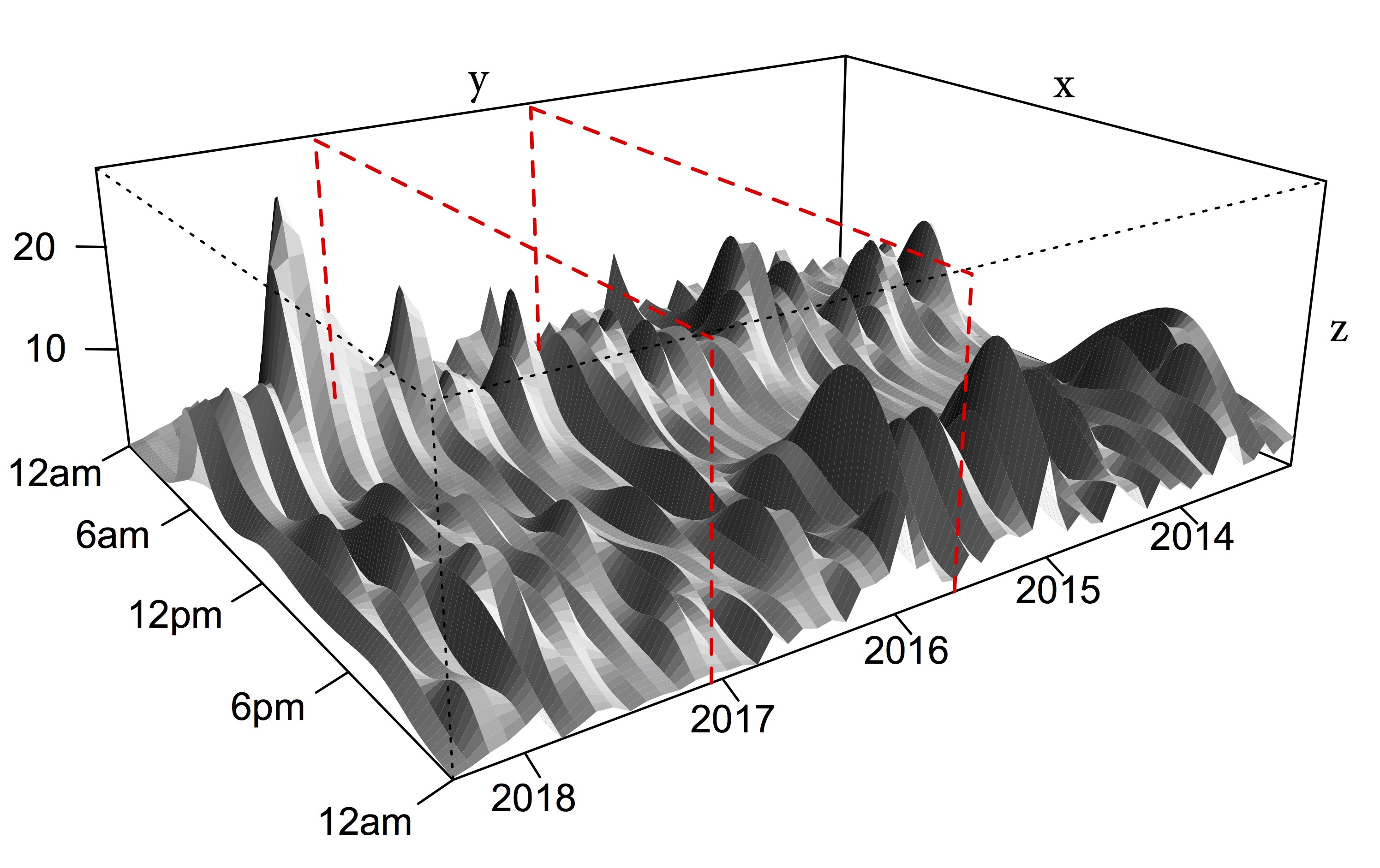}
\caption{Estimated parent hazard functions from Donald Trump's monthly Twitter data. The $x$-axis shows the time within a day, the $y$-axis shows the month and the $z$-axis shows the hazard function value. The two red dashed lines mark June 2015 (candidacy announcement) and January 2017 (assumed office), respectively.}
\label{trump2}
\end{figure}
\noindent Figure~\ref{trump1} provides some interesting insights on how Trump's tweeting behavior evolved before, during and after the presidential campaign. Here we highlight a few:
\begin{enumerate}
\item[(a)] The estimated $\alpha$ values suggested that how Trump initialized an episode of using Twitter went through notable changes over time. He started an episode about equally likely with either an original tweet or a retweet before the candidacy announcement, increasingly likely with an original tweet during the presidential campaign, and almost always with an original tweet since the presidency.
\item[(b)] Since the start of the campaign, Trump spent increasingly more time on writing each original tweet. A larger $\rho_1$ (or $\rho_0$) value indicates a higher original tweet (or retweet) rate. The estimated $\rho_1$ showed a steady decrease since the start of the campaign, suggesting that Trump spent increasingly more time on writing each original tweet. 
\item[(c)] Before Trump announced his candidacy, he posted on average 4-5 tweets per episode. This number steadily dropped during the campaign and eventually stabilized at around 2.5 since he assumed office. The number of tweets per episode is calculated using \eqref{prop1}.
\item[(d)] Trump typically spent around 15-30 minutes every time he used Twitter. This measurement of episode length remained relatively constant over time and appeared to have a slight increase since the presidency. This increase is likely attributed to the fact that, since the presidency, he had mostly original tweets in each episode and original tweets took more time to compose. The episode length is calculated using \eqref{prop2}.
\end{enumerate}

\noindent Figure~\ref{trump2} shows the estimated time-varying parent hazard function, which describes how likely Trump was to start using Twitter at any given time of the day. 
We can see that the activity level was consistently high in the morning around 6am-7am. 
The morning activity level seemed to have increased slightly since the presidency.
Before the campaign, there was high activity in the the early evening with active periods concentrated roughly around 6pm-8pm. The activity in the early evening had a noticeable decrease since the start of the campaign in June 2015, and remained low during the presidency.

To investigate the goodness of fit, we consider the procedures discussed in Section~\ref{sec::good} for each model fitted using the monthly Twitter data. The goodness-of-fit plots generally suggest that our proposed model fits the data well (see supplementary material).

\subsection{Sina Weibo data}
\label{sec::sina}
We analyze contains posting times from 5,918 Sina Weibo users. Sina Weibo, akin to a hybrid of Facebook and Twitter, is one of the most popular social media sites in China. The data were collected from followers of an official Weibo account. Restricted by the site's API policy, 5,918 of the following accounts were sampled. For each user, all posting times during the period of January 1st to January 30th, 2014 were collected. In addition, information such as the numbers of followers and followees of each user were also available. Similar to Twitter, many users on Sina Weibo are inactive users, i.e., users who do not (or very infrequently) create any content. In our study, we focus on the sampled followers who had at least 30 posts in our 30-day observation window. This subset of the sample contains 1,714 subjects. 

We fit the proposed bivariate point process model to each of the 1,714 users in the Sina Weibo data. We model the offspring gap times using \eqref{expgap} and the parent hazard function using (\ref{nonpara}) with 7 equally spaced knots in one day. Furthermore, we consider a sub-window with length $s=7$ days. To investigate the goodness of fit, we apply again the procedures discussed in Section~\ref{sec::good} for the model fitted to each user's data. The goodness-of-fit plots generally suggest that our proposed model fits the data well. We also fitted the bivariate Hawkes process in \eqref{eqn:hawkes}. The goodness-of-fit plots again suggest poor fit. The goodness-of-fit plots from both methods are included in the supplementary material.

\begin{figure}[!t]
\centering
\makebox{\includegraphics[width=0.75\textwidth]{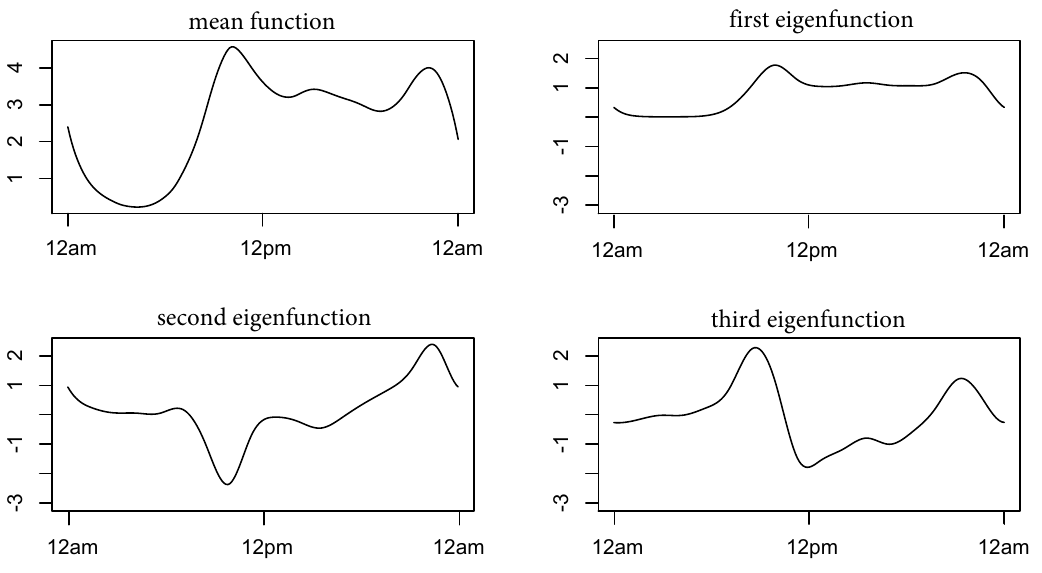}}
\caption{The mean function and the first three eigenfunctions in the functional principal component analysis of the parent hazard functions.}
\label{fig6}
\end{figure}

\bigskip
\noindent
\textbf{Characterize Sina Weibo user behavior.}
For the fitted parent hazard functions from the users, we use functional principal component analysis to investigate the dominant modes of variation. Figure~\ref{fig6} shows the mean function and the first three eigenfunctions from the analysis.
One notable pattern in the mean function is the extremely low activity level from 1am to 6am. This is expected as most users would be resting during this time. Two high activity levels appear around 9am-10am and 10pm-11pm. The first eigenfunction characterizes activeness from 8am to 12am with two moderate peaks around 10am and 10pm. The second eigenfunction describes contrasting activeness at around 10am and 10pm. This indicates that some users only had one activity peak at either 10am or 10pm. Similarly, the third eigenfunction suggests that some users were active in the morning (around 10am) and at night (around 10pm) but inactive during the time in between, while others were most active around noon but inactive in the morning and at night. These three eigenfunctions explain 76.61\% of the total variation.

\begin{figure}[!t]
\centering
\includegraphics[width=0.75\textwidth]{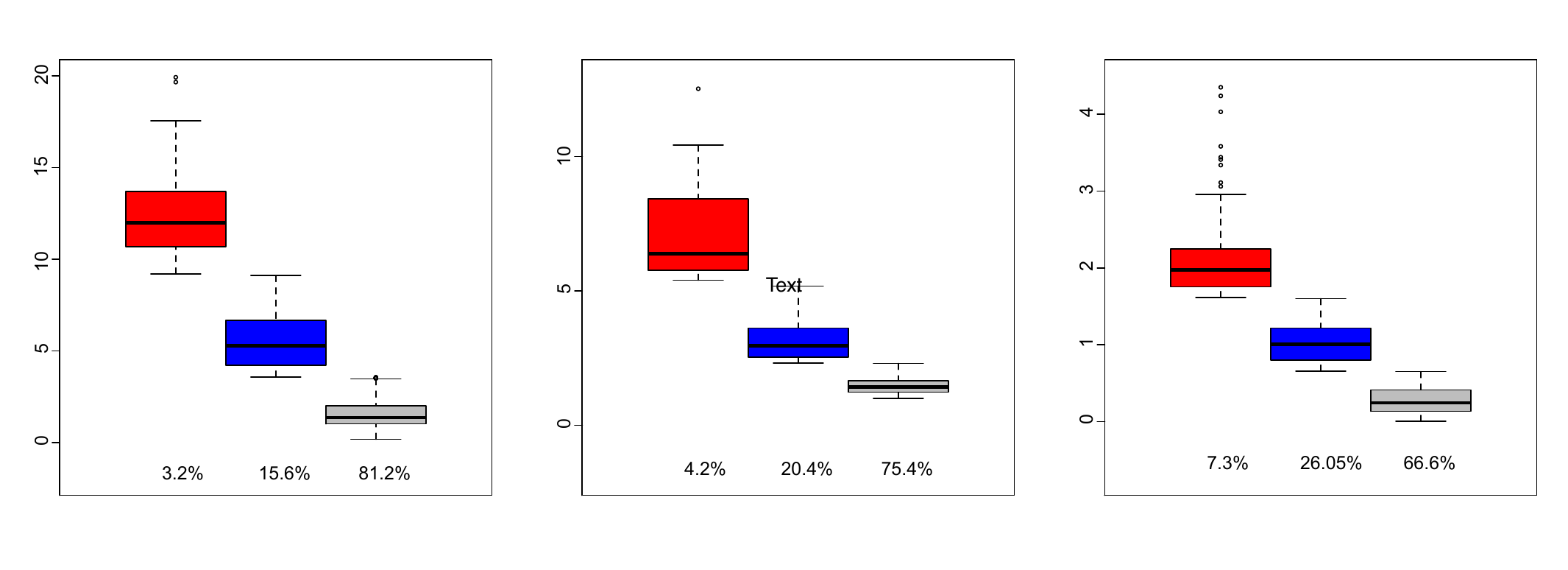}
\caption{Groups in the average daily parent hazard (left plot), average number of posts per episode (middle plot) and average length (in hours) of an episode (right plot). The percentages at the bottom of the boxplots show the percentage of users in each group.}
\label{fig7}
\end{figure}

Additionally, the parameter estimates form our model enable us to quantify the user content generating behavior in the following three measurements: (i) the average daily parent hazard function, which indicates how often a user uses Weibo; (ii) the expected number of posts per episode, which measures the activity level once a user starts using Weibo; (iii) the expected length of an episode, which measures the length of engagement once a user starts using Weibo. For each measure, K-mean clustering suggests that there are three user groups, namely high, medium and low groups. The distribution of each measure is highly skewed with the high group containing a very small percentage of users, and the low groups containing the majority of users. Figure~\ref{fig7} shows the user groups in each of the three measurements.
For the expected number of posts per episode, for example, we can see about 75\% of the users had, on average, 1.5 posts per episode; the high group had 7.5 posts per episode and it contains only 4.2\% of the users. 
For the expected length of an episode, 7\% of users (high group) had episodes that last on average 2 hours while 66.6\% of the users (low group) had episodes that lasted 16 minutes on average. The medium group, which contains about 26\% of the users, has an average episode length of 1 hour.

\bigskip
\noindent
\textbf{Social effect on users of Sina Weibo.}
For each Sina Weibo user, we also have the number of accounts the user was following, which we denote as $n_\rightarrow$, and the number of accounts that were following this user, which we denote as $n_\leftarrow$. The values for $n_\leftarrow$ are extremely skewed, ranging from 5 to 82 million (the values for $n_\rightarrow$ only range from 0 to 3000); therefore, we consider $\log(n_\leftarrow)$ in our analysis. In the following discussion, the standard error estimation provided after ``$\pm$" is calculated using bootstrap with 10,000 replications. Studying the correlations between the estimated parameters and $\log(n_\leftarrow)$, $n_\rightarrow$ leads to some interesting insights, as summarized below:
\begin{enumerate}
\item[(a)] Users who followed many accounts tended to have more reposts, as we find a correlation between $n_\rightarrow$ and $\mu_0$ ($r=0.205\pm0.035$). One explanation could be that the more accounts a user follows, the more content they can repost from. Another plausible explanation is that the ``followers" in the social media tend to repost more. 
\item[(b)] The ``popular" users, i.e., those who had many followers, tended to post more original content, as the correlation between $\log(n_\leftarrow)$ and $\mu_1$ is $0.127\pm0.026$. The ``popular" users were also more likely to initiate their Weibo engagement by posting original content, as the correlation between $\log(n_\leftarrow)$ and $\alpha$ is $0.218\pm0.023$. 
\item[(c)] The ``popular" users tended to spend more time on Sina Weibo once they started an episode of engagement, as the correlation between the expected length of an episode and $\log(n_\leftarrow)$ is $0.240\pm0.026$. Moreover, these users tended to use Sina Weibo more often, as the correlation between the average daily parent hazard and $\log(n_\leftarrow)$ is $0.242\pm0.022$.
\end{enumerate}

\section{Discussion}
\label{sec::dis}

In applications where there is only one type of events, the proposed bivariate model can be easily modified to accommodate the univariate scenario. In this case, each episode would contain only one type of events and the alternating segments within each episode no longer need to be considered. 
In the model specification, we may set $\alpha=1$, $\gamma=0$, $\mu_0$=0 and $\rho_0$=0 and estimate only $\mu_1$, $\rho_1$ and $\bbeta$. This univariate model can be considered as a special case of the proposed bivariate model.

{Our model assumes the offspring gap times to follow an exponential distribution and the number of segments in an episode as well as the number of events in a segment to follow a Poisson distribution. To evaluate whether these model assumptions are reasonable, the goodness-of-fit procedure detailed in Section B.3 can be employed. As shown in Section C of the supplementary material, the goodness-of-fit plots suggest that the fitted models are in good agreement with the observed data for both the twitter data set and the Sina Weibo data set.}

Under our proposed modeling framework, we may consider more complex model formulations, and here we discuss a few possibilities. When fitting the proposed model to Sina Weibo user data, we assume that the gap time distributions of both offspring original posts and reposts do not vary with the time of day $t$. 
A more sophisticated model can assume that these two probability densities are functions of $t$.
Similarly, we may also assume that $\gamma$, $\mu_1$ and $\mu_0$ are functions of $t$.
Such models can capture the potentially time-varying offspring generating behavior throughout the day. 
We note that this would considerably increase the number of parameters in our model and consequently make the estimation more challenging. 
To balance complexity and flexibility, such models are not further pursued in the current article.
Considering Donald Trump's Twitter data, we are interested in investigating changes in his tweeting behavior before, during and after the presidential campaign. 
To this end, we fit our proposed model to data collected for each month within the study period. 
Another approach could be, for example, to consider a varying coefficient model, in which we assume that $\alpha$, $\gamma$, $\mu_1$, $\mu_0$, $\rho_1$, $\rho_0$, $\bbeta$ are functions of day; we may fit the model using kernel smoothing technique.
This would be an interesting topic to consider for future research.

\section*{Acknowledgment}
\noindent
Zhang's research is supported by NSF DMS-2015190. 
Zhu's research is supported by the National Natural Science Foundation of China (nos. 11901105, 71991472), and the Shanghai Sailing Program for Youth Science and Technology Excellence (19YF1402700). 
Wang's research is partially supported by National Natural Science Foundation of China (No. 11831008) and the Open Research Fund of Key Laboratory of Advanced Theory and Application in Statistics and Data Science (KLATASDS-MOE-ECNU-KLATASDS2101).
Xu's research is supported by NSF SES-1902195 and Guan's research is supported by NSF SES-1758575.

\newpage
\setcounter{page}{1}
\begin{center}
{\Large\bf Supplementary Material} \\
\medskip
\end{center}

\section*{A. Proofs of main results}
\subsection*{A.1. Proof of \eqref{prop1} and \eqref{prop2} }
Here we outline the main steps in the proof of Theorem 1.
Denote the number of segments in an episode as $\Gamma+1$, where $\Gamma$ follows a Poisson distribution with mean $\gamma$. In our proposed model, the expected number of offspring in an original post segment is $1+\mu_1$, denoted as $b_1$, and the expected number of offspring in an repost segment is $1+\mu_0$, denoted as $b_0$.
The expected number of offspring in an episode can be calculated as 
\begin{eqnarray*}
&&\sum_{k=0}^{\infty}P(\Gamma=2k)(k(b_1+b_0)+\alpha b_1+(1-\alpha)b_0)+\sum_{k=0}^{\infty}P(\Gamma=2k+1)(k+1)(b_1+b_0)\\
&&=[\alpha b_1+(1-\alpha)b_0]\sum_{k=0}^{\infty}\frac{\gamma^{2k}e^{-\gamma}}{(2k)!}+\frac{b_1+b_0}{2}\sum_{k=0}^{\infty}\frac{\gamma^{2k+1}e^{-\gamma}}{(2k+1)!}+\frac{b_1+b_0}{2}\times\gamma\\
&&=[\alpha b_1+(1-\alpha)b_0]\sum_{k=0}^{\infty}\frac{\gamma^{2k}e^{-\gamma}}{(2k)!}+\frac{b_1+b_0}{2}\left(1-\sum_{k=0}^{\infty}\frac{\gamma^{2k}e^{-\gamma}}{(2k)!}\right)+\frac{b_1+b_0}{2}\times\gamma\\
&&=\frac{b_1+b_0}{2}(\gamma+1)+(\alpha-\frac{1}{2})(b_1-b_0)\sum_{k=0}^{\infty}\frac{\gamma^{2k}e^{-\gamma}}{(2k)!}.
\end{eqnarray*}
The expected length of an episode can be calculated following similar steps. Here we omit the details.

\subsection*{A.2. Proof of Theorem 1}
In the E-step, we have
\begin{eqnarray*}
Q(\btheta|\btheta')&=&\frac{1}{T}\sum_{m=1}^ME_{\Y_m|\t_m,\x_m,\btheta'}\log \left(f(\t_m,\x_m,\Y_m|\btheta)\right)\\
&=&\frac{1}{T}\sum_{m=1}^M\sum_{\y_m}\log \left(f(\t_m,\x_m,\y_m|\btheta)\right)f(\y_m|\t_m,\x_m,\btheta'),
\end{eqnarray*}
where $f(\y_m|\t_m,\x_m,\btheta')$ is the conditional distribution of $\y_m$.
Define
$$
H(\btheta|\btheta')=\frac{1}{T}\sum_{m=1}^M\sum_{\y_m}\log \left (f(\y_m|\t_m,\x_m,\btheta)\right)f(\y_m|\t_m,\x_m,\btheta').
$$
We carry out the proof in three steps.\\
\textbf{Step 1}: we want to show $Q(\btheta|\btheta')-H(\btheta|\btheta')=\ell_s^c(\btheta|\t,\x)$.\\
\begin{eqnarray*}
Q(\btheta|\btheta')-H(\btheta|\btheta')&=&\frac{1}{T}\sum_{m=1}^M\sum_{\y_m}\log\frac{f(\t_m,\x_m,\y_m|\btheta)}{f(\y_m|\t_m,\x_m,\btheta)}f(\y_m|\t_m,\x_m,\btheta')\\
&=&\frac{1}{T}\sum_{m=1}^M\sum_{\y_m}\log\left(f(\t_m,\x_m|\btheta)\right)f(\y_m|\t_m,\x_m,\btheta')\\
&=&\frac{1}{T}\sum_{m=1}^M\log\left(f(\t_m,\x_m|\btheta)\right)=\ell_s^c(\btheta|\t,\x)
\end{eqnarray*}
\textbf{Step 2}: we want to show $H(\btheta'|\btheta)\le H(\btheta|\btheta)$.\\
We have
$$
H(\btheta'|\btheta)=\frac{1}{T}\sum_{m=1}^M\sum_{\y_m}\log \left (f(\y_m|\t_m,\x_m,\btheta')\right)f(\y_m|\t_m,\x_m,\btheta)
$$
and
$$
H(\btheta|\btheta)=\frac{1}{T}\sum_{m=1}^M\sum_{\y_m}\log \left (f(\y_m|\t_m,\x_m,\btheta)\right)f(\y_m|\t_m,\x_m,\btheta).
$$
Therefore,
$$
H(\btheta'|\btheta)-H(\btheta|\btheta)=\frac{1}{T}\sum_{m=1}^M\sum_{\y_m}\log \frac{f(\y_m|\t_m,\x_m,\btheta')}{f(\y_m|\t_m,\x_m,\btheta)}f(\y_m|\t_m,\x_m,\btheta)\le 0.
$$
\textbf{Step 3}: we want to show $\ell_s^c(\btheta_{p}|\t,\x)\ge\ell_s^c(\btheta_{p-1}|\t,\x)$, where $\btheta_{p}=\arg\max_{\btheta}Q(\btheta|\btheta_{p-1})$.
This is true because
\begin{eqnarray*}
\ell_s^c(\btheta_{p}|\t,\x)&=&Q(\btheta_{p}|\btheta_{p-1})-H(\btheta_{p}|\btheta_{p-1})\\
&\ge&Q(\btheta_{p-1}|\btheta_{p-1})-H(\btheta_{p-1}|\btheta_{p-1})=\ell_s^c(\btheta_{p-1}|\t,\x).
\end{eqnarray*}
The inequality established from the results in Step 2 and the fact that $\btheta_{p}$ maximizes $Q(\btheta|\btheta_{p-1})$.

\subsection*{A.3. Proof of Theorem 2}
We first give the following lemma from Theorem 3.1 in Crowder (1986).
\begin{lemma} Define $\hat\btheta_{M,s}$ as the solution to $U_{M,s}(\btheta)=0$. Let $B_{\epsilon}=\{|\btheta-\btheta_s^*|<\epsilon\}$ for some $\epsilon>0$. Then $P[\hat\btheta_{M,s}\in B_{\epsilon}]\rightarrow 1$ if
\begin{description}
\item{(A1)} $E[U_{M,s}(\btheta)]\rightarrow0$ only at $\btheta_s^*$,
\item{(A2)} $\inf_{\bm{\Theta}-B_{\epsilon}}|E[U_{M,s}(\btheta)]|\ge C_{\epsilon}$ for some $C_{\epsilon}>0$,
\item{(A3)} $\sup_{\bm{\Theta}-B_{\epsilon}}|U_{M,s}(\btheta)-E[U_{M,s}(\btheta)]|\rightarrow0$ in probability.
\end{description}
\end{lemma}
To show that $\hat\btheta_{M,s}\rightarrow\btheta_s^*$, we need to verify conditions (A1)-(A3) for $U_{M,s}(\btheta)$. Since $E[U_{M,s}(\btheta)]=0$ only when $\btheta=\btheta_s^*$, to verify (A1) and (A2), it is sufficient to show that
\begin{description}
\item{(C1)} $E[U_{M,s}(\btheta)]$ is continuous in $\btheta$.
\end{description}
Since $\bm{\Theta}$ is assumed to be a compact set, if (C1) is true, then we have $E[U_{M,s}(\btheta)]$ is bounded over $\btheta\in\bm{\Theta}$.
To verify (A3), it is sufficient to show that for any $\epsilon>0$, $\eta>0$, there exist $\delta>0$ and $M'<\infty$ such that the following two conditions are satisfied for $M>M'$(Guan, 2006):
\begin{description}
\item{(C1*)} $\sup_{|\btheta-\btheta_s^*|<\delta}|E[U_{M,s}(\btheta_1)]-E[U_{M,s}(\btheta_2)]|<\epsilon/2$,
\item{(C2)} $P[\sup_{|\btheta-\btheta_s^*|<\delta}|U_{M,s}(\btheta_1)-U_{M,s}(\btheta_2)|>\epsilon/2]<\eta$.
\end{description}
(C1*) directly follows from (C1) and the fact that $\bm\Theta$ is compact. 
Therefore, to show $\hat\btheta_{M,s}\rightarrow\btheta_s^*$, we only need to verify (C1) and (C2).
To show (C1), it is sufficient to show that $E[U^\prime_{M,s}(\btheta)]$ is bounded.
We have
$$
U^\prime_{M,s}(\btheta)=\frac{1}{T}\sum_{m=1}^M\left[\frac{f^{(2)}(\t_m,\x_m|\btheta)}{f(\t_m,\x_m|\btheta)}-\left(\frac{f^{(1)}(\t_m,\x_m|\btheta)}{f(\t_m,\x_m|\btheta)}\right)^2\right].
$$
If there exists a nonnegative function $\kappa(\cdot)$ such that $E[\kappa(|\t_m|)^2]<\infty$ and 
$$
\left|\frac{f^{(1)}(\t_m,\x_m|\btheta)}{f(\t_m,\x_m|\btheta)}\right|<\kappa(|\t_m|),\quad \left|\frac{f^{(2)}(\t_m,\x_m|\btheta)}{f(\t_m,\x_m|\btheta)}\right|<\kappa(|\t_m|),
$$
then we have
\begin{eqnarray*}
|E[U^\prime_{M,s}(\btheta)]|&\le& \frac{1}{T}\sum_{m=1}^ME\left|\frac{f^{(2)}(\t_m,\x_m|\btheta)}{f(\t_m,\x_m|\btheta)}\right|+\frac{1}{T}\sum_{m=1}^ME\left(\frac{f^{(1)}(\t_m,\x_m|\btheta)}{f(\t_m,\x_m|\btheta)}\right)^2\\
&\le&\frac{1}{T}\sum_{m=1}^ME[\kappa(|\t_m|)+\kappa(|\t_m|)^2]<\infty.
\end{eqnarray*}
Moreover, by the continuity of $U_{M,s}(\btheta)$, we have
$$
\sup_{|\btheta-\btheta_s^*|<\delta}|U_{M,s}(\btheta_1)-U_{M,s}(\btheta_2)|<\delta^2\sup_{\btheta}|U'_{M,s}(\btheta)|.
$$
Therefore, to show (C2), it is sufficient to show that for any $\eta>0$, there exists $M_\eta<\infty$ such that
$$
P[\sup_{\btheta}|U_{M,s}'(\btheta)|>M_\eta]<\eta.
$$
We have
\begin{eqnarray*}
|U^\prime_{M,s}(\btheta)|&\le&\frac{1}{T}\sum_{m=1}^M\left|\frac{f^{(2)}(\t_m,\x_m|\btheta)}{f(\t_m,\x_m|\btheta)}\right|+\frac{1}{T}\sum_{m=1}^M\left(\frac{f^{(1)}(\t_m,\x_m|\btheta)}{f(\t_m,\x_m|\btheta)}\right)^2\\
&\le& \frac{1}{T}\sum_{m=1}^M[\kappa(|\t_m|)+\kappa(|\t_m|)^2].
\end{eqnarray*}
Since we assume that $\kappa(\cdot)$ is a nonnegative function and $E[\kappa(|\t_m|)^2]$ is bounded, there exists $m_0<\infty$ such that $E[\kappa(|\t_m|)+\kappa(|\t_m|)^2]<m_0$.
By the Markov inequality, we have
$$
P\{\kappa(|\t_m|)+\kappa(|\t_m|)^2>m_0/\eta\}<\eta.
$$
Write $M_\eta=m_0/\eta$, we have
$$
P[\sup_{\btheta}|U'_{M,s}(\btheta)|>M_\eta]\le P\{\kappa(|\t_m|)+\kappa(|\t_m|)^2>M_\eta\}<\eta.
$$
Thus, we have shown (C2). 

Next, we want to show that $\forall\epsilon>0$, there exists $s'\in\mathcal{R}$ such that $\forall$ $s>s_0$, we have
$$
||\btheta_s^*-\btheta_0||_{\infty}<\epsilon.
$$
If the above statement is not true, then there exists $\epsilon^*\in\mathcal{R}$, such that $\forall s>0$, there exists $\tilde{s}>s$,
such that $||\btheta_{\tilde s}^*-\btheta_0||_{\infty}>\epsilon^*$.
Therefore, there exists a sequence $\{\btheta^*_{s_n}\}_{n=1}^{\infty}$ such that $s_n\rightarrow\infty$ and $||\btheta_{s_n}^*-\btheta_0||_{\infty}>\epsilon^*$.
Since $\bm{\Theta}$ is compact, there exists a subsequence $\{\btheta^*_{s_{n_k}}\}_{k=1}^{\infty}$ such that
$\btheta^*_{s_{n_k}}\rightarrow\btheta_0^*$ as $k\rightarrow\infty$ and $||\btheta_0^*-\btheta_0||_{\infty}>\epsilon^*$.
Define $\Psi_s(\btheta)=E[U_{m,s}(\btheta)]$.
We have
$$
|\Psi_{s_{n_k}}(\btheta^*_{s_{n_k}})-\Psi_{s_{n_k}}(\btheta_0^*)|=|\Psi^\prime_{s_{n_k}}(\tilde\btheta_0)||\btheta_{s_{n_k}}^*-\btheta_0^*|.
$$
By definition, we have $\Psi_{s_{n_k}}(\btheta^*_{s_{n_k}})=0$.
Since $\btheta^*_{s_{n_k}}\rightarrow\btheta_0^*$ and $\Psi^\prime_s(\btheta)$ is bounded and continuous (from the proof of Theorem 3),
we have $\Psi_{s_{n_k}}(\btheta_0^*)\rightarrow 0$ as $k\rightarrow\infty$. This contradicts with the assumption that $\Psi_{s}(\btheta)\rightarrow 0$ as $s\rightarrow\infty$ only at $\btheta=\btheta_0$.

\subsection*{A.4. Proof of Lemma 1}
To verify condition (3.1) and (3.2) in Theorem 3, we show that there exists a nonnegative function $\kappa(\cdot)$ such that $E[\kappa(|\t_m|)^2]<\infty$ and 
$$
\left|\frac{f^{(1)}(\t,\x|\btheta)}{f(\t,\x|\btheta)}\right|<\kappa(|\t|),\quad \left|\frac{f^{(2)}(\t,\x|\btheta)}{f(\t,\x|\btheta)}\right|<\kappa(|\t|),
$$
where $f(\t,\x|\btheta)=\sum_{\y}f(\t,\y,\x|\btheta)$. 

Define 
$$
h_{1}(\t,\x,\y|\alpha)=\prod_{l=1}^n\alpha^{I(y_l=1,x_l=1)}(1-\alpha)^{I(y_l=1, x_l=0)},
$$
$$
h_{2}(\t,\x,\y|\gamma)=\prod_{k=1}^K{\gamma^{n_k-1}e^{-\gamma}},
$$
$$
h_{3}(\t,\x,\y|\mu_1)=\prod_{k=1}^K\prod_{j=1}^{n_k}(\mu_1^{l_{k_j}-1}e^{-\mu_1})^{I(z_{k_j}=1)},
$$
$$
h_{4}(\t,\x,\y|\mu_0)=\prod_{k=1}^K\prod_{j=1}^{n_k}(\mu_0^{l_{k_j}-1}e^{-\mu_0})^{I(z_{k_j}=0)}.
$$
$$
h_{5}(\t,\x,\y|\rho_1)=\prod_{l=1}^n[\rho_1\exp(-\rho_1(t_l-t_{l-1})]^{I(y_l=0, x_l=1)},
$$
$$
h_{6}(\t,\x,\y|\rho_0)=\prod_{l=1}^n[\rho_0\exp(-\rho_0(t_l-t_{l-1})]^{I(y_l=0, x_l=0)},
$$
$$
h_{7}(\t,\x,\y|\bbeta)=\prod_{l=1}^n \left[\lambda(t_l;\bbeta)\exp\left(-\int_{t_{l-1}}^{t_l} \lambda(t;\bbeta)\dd t\right)\right]^{I(y_l=1)}\times\exp\left(-\int_{t_n}^{T} \lambda(t;\bbeta)\dd t\right).
$$
Then we can write (\ref{full}) as
$$
f(\t,\x,\y|\btheta)=C(\x,\y)\prod_{i=1}^7h_i(\t,\x,\y|\btheta),
$$
where $C(\x,\y)$ is a function of $\x$ and $\y$ only. Note that $h_1(\t,\x,\y|\btheta),\ldots, h_6(\t,\x,\y|\btheta)$ are all non-negative functions.

We have
\begin{eqnarray*}
\left|\frac{\dot{h}_{1}(\t,\x,\y|\alpha)}{h_{1}(\t,\x,\y|\alpha)}\right|&=&\left|\frac{\sum_{l=1}^nI(y_l=1,x_l=1)}{\alpha}-\frac{\sum_{l=1}^nI(y_l=1,x_l=0)}{1-\alpha}\right|\\
&\le&n\left|\frac{1}{\alpha}+\frac{1}{1-\alpha}\right|.
\end{eqnarray*}
Furthermore, since
\begin{eqnarray*}
\frac{\ddot{h}_{1}(\t,\x,\y|\alpha)}{h_{1}(\t,\x,\y|\alpha)}&=&\left(\frac{\sum_{l=1}^nI(y_l=1,x_l=1)}{\alpha}-\frac{\sum_{l=1}^nI(y_l=1,x_l=0)}{1-\alpha}\right)^2\\
&&-\frac{\sum_{l=1}^nI(y_l=1,x_l=1)}{\alpha^2}-\frac{\sum_{l=1}^nI(y_l=1,x_l=0)}{(1-\alpha)^2},
\end{eqnarray*}
we have
\begin{eqnarray*}
\left|\frac{\ddot{h}_{1}(\t,\x,\y|\alpha)}{h_{1}(\t,\x,\y|\alpha)}\right|&\le&n^2\left(\frac{1}{\alpha}+\frac{1}{1-\alpha}\right)^2+n\left|\frac{1}{\alpha^2}+\frac{1}{(1-\alpha)^2}\right|.
\end{eqnarray*}
Since $\bm\Theta$ is assumed to be compact, we can find $\kappa_1(\cdot)$ such that
$$
\left|\frac{\dot{h}_{1}(\t,\x,\y|\alpha)}{h_{1}(\t,\x,\y|\alpha)}\right|\le\kappa_1(n), \quad \left|\frac{\ddot{h}_{1}(\t,\x,\y|\alpha)}{h_{1}(\t,\x,\y|\alpha)}\right|\le\kappa_1(n).
$$
Hence, we have
\begin{eqnarray*}
\frac{\partial f(\t,\x|\btheta)}{\partial\alpha}&=&\sum_{\y}\frac{\partial f(\t,\x,\y|\btheta)}{\partial\alpha}
=\sum_{\y}C(\x,\y) \prod_{i=2}^7h_i(\t,\x,\y|\btheta)\frac{\partial{h_1}(\t,\x,\y|\alpha)}{\partial\alpha}\\
&\le&\kappa_1(n)\sum_{\y}f(\t,\x,\y|\btheta)=\kappa_1(n)f(\t,\x|\btheta),
\end{eqnarray*}
and
\begin{eqnarray*}
\frac{\partial^2 f(\t,\x|\btheta)}{\partial\alpha^2}&=&\sum_{\y}\frac{\partial^2 f(\t,\x,\y|\btheta)}{\partial\alpha^2}
=\sum_{\y}C(\x,\y) \prod_{i=2}^7h_i(\t,\x,\y|\btheta)\frac{\partial^2{h_1}(\t,\x,\y|\alpha)}{\partial\alpha^2}\\
&\le&\kappa_1(n)\sum_{\y}f(\t,\x,\y|\btheta)=\kappa_1(n)f(\t,\x|\btheta).
\end{eqnarray*}
Following similar steps, we can show that there exists $\kappa_j(\cdot)$, $ j=2,\ldots,7$ such that
$$
\frac{\partial f(\t,\x|\btheta)}{\partial\theta_j}\le \kappa_j(n)f(\t,\x|\btheta),
$$
$$
\frac{\partial^2 f(\t,\x|\btheta)}{\partial\theta_j^2}\le \kappa_j(n)f(\t,\x|\btheta),\quad j=2,\ldots,7,
$$
and $\kappa_j(n)=O(n^2)$, $j=2,\ldots,7$.
Define $\kappa(.)=\max_{j=1}^7 \kappa_j(.)$. We have that 
$$
\left|\frac{f^{(1)}(\t,\x|\btheta)}{f(\t,\x|\btheta)}\right|\le\kappa(n), \quad \left|\frac{f^{(2)}(\t,\x|\btheta)}{f(\t,\x|\btheta)}\right|\le\kappa(n).
$$
Now it remains for us to show that $E[\kappa(|\t_m|)^2]<\infty$.
Since $\kappa_j(n)=O(n^2)$, $j=1,\ldots,7$, 
it is sufficient for us to show that $E|\t_m|^{4+\epsilon}$ is finite, for some $\epsilon>0$.
This is true by observing that the conditional intensity of our proposed point process is always smaller than $\rho_{\max}=\max\{\rho_1,\rho_0,\max_{t\in[0,1]}\lambda(t,\bbeta)\}$. 
Therefore $E|\t_m|^{4+\epsilon}\le E[n_{\max}]^{4+\epsilon}<\infty$, where $n_{\max}$ is the number of events in a sub-window from a homogeneous Poisson process with rate $\rho_{\max}$.


\subsection*{A.5. Proof of Theorem 3}
{The proof is divided into two steps. In step 1, we show that 
$\sqrt{T}(\hat\btheta_{M,s}-\btheta^*_s)$ converges in distribution to $N(\bm{0},\mathcal{I}'_M{}^{-1})$, where $$
\mathcal{I}'_M{}^{-1}=\mathcal{H}^{-1}(\btheta^*_s)\left\{\frac{1}{T}\sum_{i=1}^M\sum_{j=1}^M \mathbb{E}\left[(U_{i,s}(\btheta^*_s)U^T_{j,s}(\btheta^*_s)\right]\right\}\mathcal{H}^{-1}(\btheta^*_s)^{T}.
$$
In step 2, we show that $\|\btheta_s^\ast-\btheta_0\|_2=o(T^{-1/2})$. 
Putting steps 1 and 2 together and by Slutsky's theorem, we arrive at the desired conclusion in Theorem 3.}

\bigskip\noindent
{
\textbf{Step 1.}
Following \cite{Bolthausen1982}, we define the following mixing coefficients to quantify the dependence in the proposed point process. Let $\mathbb{N}$ denote the set of all natural numbers. For $\Lambda\subseteq\mathbb{N}$, let $\mathcal{F}(\Lambda)$ denote the $\sigma$-algebra generated by $\Lambda$. Let $d(\Lambda_1,\Lambda_2)=\inf\{|m_1-m_2|:m_1\in\Lambda_1,m_2\in\Lambda_2\}$.
For all $v\in\mathbb{N}$ and $k,l\in\mathbb{N}\cup\{\infty\}$, define the following mixing coefficient: 
$$
\alpha_{k,l}(v)=\sup\{|P(A_1\cap A_2)-P(A_1)P(A_2)|:A_i\in\mathcal{F}(\Lambda_i),|\Lambda_1|\le k, |\Lambda_2|\le l,d(\Lambda_1,\Lambda_2)\ge v\}.
$$
The normality result in step 1 follows from \cite{Bolthausen1982}. To apply \cite{Bolthausen1982}, we need to verify the following mixing coefficient conditions. 
\begin{itemize}
\item{(i)} $\sum_{v=1}^\infty\alpha_{k,l}(v)<\infty$ for $k+l\le 4$, 
\item{(ii)} $\alpha_{1,\infty}(v)=o(v^{-1})$, 
\item{(iii)} $\mathbb{E}[(U_{m,s}(\btheta^*_s))^{2+\delta}]<\infty$ and $\sum_{v=1}^\infty\alpha_{1,1}(v)^{\delta/(2+\delta)}<\infty$ for some $\delta>0$. 
\end{itemize}
Next, we verify the above mixing coefficient conditions.}

{For $v\in\mathbb{N}$ and $k,l\in\mathbb{N}\cup\{\infty\}$, consider the mixing coefficient $\alpha_{k,l}(v)=\sup\{|P(A_1\cap A_2)-P(A_1)P(A_2)|:A_i\in\mathcal{F}(\Lambda_i),|\Lambda_1|\le k, |\Lambda_2|\le l,d(\Lambda_1,\Lambda_2)\ge v\}.$ In the following proof, without loss of generality, we assume $\inf_{m_1\in\Lambda_1,m_2\in\Lambda_2} m_2-m_1>0$, i.e.,
elements in $\Lambda_1$ are smaller than those in $\Lambda_2$. 
Let $t_1$ be the last event of the last episode started in $\Lambda_1$ and $t_2$ be the parent of the first event in $\Lambda_2$. Define $B$ as the event that all episodes in $\Lambda_1$ end before $\Lambda_2$, i.e., $B=\{t_2>t_1\}$. 
Given $t_1$, we have
\begin{eqnarray*}
P(A_1\cap A_2)&=&P(A_1\cap A_2\cap B)+P(A_1\cap A_2\cap B^c)\\
&=&P(A_1\cap A_2|B)P(B)+P(A_1\cap A_2\cap B^c)\\
&=&P(A_1|B)P(A_2|B)P(B)+P(A_1\cap A_2\cap B^c),
\end{eqnarray*}
where the last equality holds as $A_1$ and $A_2$ are independent given $t_1$ and $t_1<t_2$. 
Next,
\begin{equation*}
\begin{aligned}
P(A_1)P(A_2)=&\{P(A_1\cap B)+P(A_1\cap B^c)\}\{P(A_2\cap B)+P(A_2\cap B^c)\}\\
=&P(A_1\cap B)P(A_2\cap B)+P(A_1\cap B)P(A_2\cap B^c)\\
&\hspace{0.8in}+P(A_1\cap B^c)\{P(A_2\cap B)+P(A_2\cap B^c)\}\\
=&P(A_1|B)P(A_2|B)P(B)P(B)+P(A_1\cap B)P(A_2\cap B^c)+P(A_1\cap B^c)P(A_2)
\end{aligned}    
\end{equation*}
Let $V$ be the length of an episode. Combining the above results, we can get that
\begin{equation*}
\begin{aligned}
&|P(A_1\cap A_2)-P(A_1)P(A_2)|\\
=&|P(A_1|B)P(A_2|B)P(B)+P(A_1\cap A_2\cap B^c)-P(A_1|B)P(A_2|B)P(B)P(B)\\
&\hspace{0.8in}-P(A_1\cap B)P(A_2\cap B^c)-P(A_1\cap B^c)P(A_2)|\\
=&|P(A_1|B)P(A_2|B)P(B)P(B^c)+P(A_1\cap A_2\cap B^c)-P(A_1\cap B)P(A_2\cap B^c)-P(A_1\cap B^c)|\\
\leq& 4P(B^c)\leq 4P(V\geq v).
\end{aligned}
\end{equation*}}

{Next, we investigate the tail behavior of $V$. We first consider the univariate case. Suppose that the offspring gap times follow an exponential distribution with parameter $\rho$. Let $W$ denote the Poisson number (with parameter $\mu$) of gap times. Note that $V|W=w\sim\hbox{Erlang}(\rho,w)$. Then, we have that
\begin{eqnarray*}
P(V>v)&=&\sum_{w=1}^\infty P(V>v|W=w)P(W=w)\\
&=&\sum_{w=1}^\infty \sum_{k=0}^{w-1}\frac{1}{k!}(\rho{vs})^k e^{-\rho{vs}} \left[\frac{\mu^w e^{-\mu}}{w!}\right].
\end{eqnarray*}
It is seen that $V$ follows a compound Poisson exponential distribution and is log-concave with an exponential rate of decay (see Theorem 5.2 in \citet{ninh2013log} and Section 5.1 in \citet{saumard2014log}). As a result, it holds that $\sum_{v=1}^\infty\alpha_{k,l}(v)<\infty$ for any given $k$ and $l$, $\alpha_{1,\infty}(v)=o(v^{-1})$ and $\sum_{v=1}^\infty\alpha_{1,1}(v)^{\delta/(2+\delta)}<\infty$ for any $\delta>0$.
}

{In the bivariate case, let $V=V_0+V_1$, where $V_0$ and $V_1$ are the sums of gap times due to repost and post events, respectively. Let $W_0$ and $W_1$ be the total number of gap times (not Poisson any more) for each type. Note that
$$
P(V>v)\le P(V_0>v/2)+P(V_1>v/2).
$$
Based on the additive property of Poisson distributions, we have
\begin{equation*}
\begin{aligned}
    P(V_i> v/2)&=\sum_{w_i=1}^\infty P(V_i> v/2| W_i=w_i)P(W_i=w_i)\\
    &\le\sum_{w_i=1}^\infty\sum_{k=0}^{w_i-1}\frac{1}{k!}(\rho_i vs/2)^k e^{-\rho_i vs/2} \sum_{j=1}^{w_i}{\frac{(j\mu_i)^{w_i-j} e^{-j \mu_i}}{(w_i-j)!}\cdot\frac{\gamma^{j-1} e^{-\gamma}}{(j-1)!}}.
\end{aligned}    
\end{equation*}
Unlike the univariate case, the event number $W_i$ in the above equation follows a compound Poisson distribution, i.e., sum of Poisson number of Poisson random variables. By Theorem 4.1 in \citet{ninh2013log}, $W_i$ has a log-concave distribution, $i=1,2$.
As $V_i\mid W_i=w_i$ is a sum of exponential random variables and $W_i$ is log-concave, by Theorem 5.2 in \citet{ninh2013log}, $V_i$ has a log-concave distribution and hence has an exponential rate of decay (see Section 5.1 in \citet{saumard2014log}). As a result, it holds that $\sum_{v=1}^\infty\alpha_{k,l}(v)<\infty$ for any given $k$ and $l$, $\alpha_{1,\infty}(v)=o(v^{-1})$ and $\sum_{v=1}^\infty\alpha_{1,1}(v)^{\delta/(2+\delta)}<\infty$ for any $\delta>0$. 
Finally, we can take $\delta=1$ and show $\mathbb{E}[(U_{m,s}(\btheta^*_s))^3]<\infty$ following the same arguments as in Section A.4 (detailed steps omitted).
Thus, we have completed step 1.}


\bigskip\noindent
{
\textbf{Step 2.} In this step, we show that $\|\btheta_s^\ast-\btheta_0\|_2=o(T^{-1/2})$.}
{To ease notation, let's first consider the univariate case. 
Consider the true joint density of $\t,\y$ on $[0,T]$ for a given $\btheta$, denoted as $f_0(\t,\y|\btheta)$. We first derive a generic result for $f_0(\t,\y|\btheta)$ on $[0,T]$ and then apply it to each sub-window considered in the composite likelihood estimation.}

{Let $k=\sum_{l=1}^n I(y_l=1)$ denote the total number of episodes.
The exact joint density function $f_0(\t,\y|\btheta)$ can be written as
\begin{eqnarray*}
f_0(\t,\y|\btheta)&=&h_1(\t,\y)\times\left[\prod_{l=2}^n \prod_{h=0}^{1}f_{lh}(d_l;\btheta)^{I(y_l=h)}\right]\prod_{i=2}^{k-1}P(N_i=n_i)\times h_2(\t,\y),
\end{eqnarray*}
where $N_i$ follows a Poisson distribution with parameter $\mu$ and $n_i$ is the number of events in the $i$th episode in $\y$. 
When $t_1$ is not a parent event, the offspring events before the first parent in $\t$ can be written as $t_1,\ldots,t_{n_1}$. Correspondingly, we have $y_1=\ldots=y_{n_1}=0$ and $y_{n_1+1}=1$.
We can then write
\begin{eqnarray*}
h_1(\t,\y)&=&{I(y_1=1)\underbrace{\lambda(t_1,\bbeta)\exp[-\int_0^{t_1}\lambda(t,\bbeta)\dd t]P(N_1=n_1)}_{h_1'(\t,\y)}/\lambda(t_1,\bbeta)}\\
&&+I(y_1=0)\exp(-\rho t_1)P(N_1>n_1),
\end{eqnarray*}
Furthermore, we have
\[
h_2(\t,\y)={\underbrace{P(N_k=n_k)\exp[-\int_{t_n}^{T}\lambda(t,\bbeta)\dd t]}_{h_2'(\t,\y)}}+P(N_k> n_k)\exp[-\rho(T-t_n)].
\]
Letting $f_{11}(d_1)=\exp[-\int_0^{t_1}\lambda(t,\bbeta)\dd t]$ and $f_{10}(d_1)=\exp(-\rho t_1)$, and we have
\begin{eqnarray}\label{eqn:f0}
&&f_0(\t,\y|\btheta)=h_1(\t,\y)\prod_{l=2}^n\prod_{h=0}^1 \left[f_{lh}(d_l;\btheta)^{I(y_l=h)}\right]\left[\prod_{i=2}^{k}{\exp(-\mu)\mu^{n_i}\over n_i!}\right]h_2(\t,\y)\\\nonumber
&=&\prod_{l=1}^n\prod_{h=0}^1 \left[f_{lh}(d_l;\btheta)^{I(y_l=h)}\right]
\left[\frac{\exp(-\mu)\mu^{n_1}}{n_1!}\right]^{I(y_1=1)}\left[1-\frac{\Gamma(n_1+1,\mu)}{n_1!}\right]^{I(y_1=0)}\prod_{i=2}^{k}{\exp(-\mu)\mu^{n_i}\over n_i!}\\\nonumber
&&\left\{\frac{\exp(-\mu)\mu^{n_k}}{n_k!}\exp[-\int_{t_n}^{T}\lambda(t,\bbeta)\dd t]+\left[1-\frac{\Gamma(n_k+1,\mu)}{n_k!}\right]\exp[-\rho(T-t_n)]\right\},
\end{eqnarray}
where $\Gamma(x,y)$ is the upper incomplete gamma function. 
The exact likelihood function $\sum_\y f_0(\t,\y|\btheta)$ calculated using the $f_0(\t,\y|\btheta)$ in the above equation takes a very complicated form.
In our calculation, we use $h_1'(\t,\y)$, $h_2'(\t,\y)$ to approximate $h_1(\t,\y)$ and $h_2(\t,\y)$ respectively, in which case, the logarithm of the approximated density $f(\t,\y|\btheta)$ can be written compactly as
\begin{eqnarray*}
\log f(\t,\y|\btheta)&=&\sum_{l=1}^n\sum_{m=0}^{1}I(y_l=m)\log f_{lm}(d_l;\btheta)+n\log\mu-\nonumber\\
&&\sum_{l=1}^n I(y_l=1) (\log\mu+\mu)-\int_{t_n}^{T} \lambda(t;\bbeta)\dd t-\sum_{i=1}^k \log n_i!.
\end{eqnarray*}
Then, it holds that 
\begin{equation*}
    \begin{aligned}
    &\log f(\t,\y|\btheta)-\log f_0(\t,\y|\btheta)=\log\left[\frac{h_1'(\t,\y)}{h_1(\t,\y)}\right]+\log\left[\frac{h_2'(\t,\y)}{h_2(\t,\y)}\right].
    \end{aligned}
\end{equation*}
Next, we move to show that it holds with probability at least $1-2c_0z^{-\rho/2}$ that
\begin{equation}
\label{eqn:bias}
\sup_{\y\in\mathcal{Y}}\left[\log f(\t,\y|\btheta)-\log f_0(\t,\y|\btheta)\right]=o((\log z)^2),
\end{equation}
where $c_0$ is a positive constant, $\mathcal{Y}$ is as defined in \eqref{like} and $z>0$ is a divergent scalar, the rate of which is to be defined later.
}

{
By the definition of $h_1(\t,\y)$ and $h_1'(\t,\y)$, we have
\begin{equation*}
    \frac{h_1'(\t,\y)}{h_1(\t,\y)}=\frac{1}{I(y_1=1)/\lambda(t_1,\bbeta)+a_0I(y_1=0)\exp\left\{-\int_0^{t_1}(\rho-\lambda(t,\bbeta))\dd t\right\}},
\end{equation*}
where $a_0=\frac{P(N_1>n_1)}{P(N_1=n_1)}$. If $y_1=1$, $\log\left[\frac{h_1'(\t,\y)}{h_1(\t,\y)}\right]=\log\lambda(t_1,\bbeta)$. If $y_1=0$, we have
\begin{equation*}
    \log\left[\frac{h_1'(\t,\y)}{h_1(\t,\y)}\right]=\log(1/a_0)\int_0^{t_1}(\rho-\lambda(t,\bbeta))\dd t.
\end{equation*}
As the parameter space is assumed to be compact, we have $|\int_0^{t_1}(\rho-\lambda(t,\bbeta))\dd t|\le C_1t_1$ for some positive constant $C_1$.
For $a_0$, we have
\begin{equation*}
    a_0=\frac{\sum_{i=1}^\infty\frac{e^{-\mu}\mu^{n_1+1}}{(n_1+i)!}}{\frac{e^{-\mu}\mu^{n_1}}{n_1!}}=\sum_{i=1}^\infty\frac{\mu^i n_1!}{(n_1+i)!}\leq\sum_{i=1}^\infty\frac{\mu^i }{i!}=e^{\mu}-1,
\end{equation*} and
$a_0\geq \frac{P(N_1=n_1+1)}{P(N_1=n_1)}=\frac{\mu}{n_1+1}$. Since $N_1$ follows a Poisson distribution, using the Chernoff inequality, we have $P(n_1\geq z)\leq \exp^{-\frac{(z-\mu)^2}{z}}$. The lower and upper bounds of $a_0$ together imply that $\log(1/a_0)=o(\log z)$ with probability at least $1-\exp^{-\frac{(z-\mu)^2}{z}}$. We also have that $P(t_1>\log z)\leq \exp(-\rho\log z/2)$. 
Combining the above arguments, we can get that $\log\left[\frac{h_1'(\t,\y)}{h_1(\t,\y)}\right]=o((\log z)^2)$ with probability at least $1-c_0z^{-\rho/2}$ for some positive constant $c_0$. 
Similarly, let $b_0=\frac{P(N_k>n_k)}{P(N_k=n_k)}$, we have
\begin{equation*}
    \frac{h_2'(\t,\y)}{h_2(\t,\y)}=\frac{1}{1+b_0\exp\left\{-\int_{t_n}^{T}(\rho-\lambda(t,\bbeta))\dd t\right\}}.
\end{equation*}
As the parameter space is assumed to be compact, we have 
$1\le 1+b_0\exp\left\{-\int_{t_n}^{T}(\rho-\lambda(t,\bbeta))\right\}\le 1+C_2b_0(T-t_n)$ for some positive constant $C_2$. 
Similar to $a_0$, we can show that $b_0\le e^\mu-1$. Moreover, when $P(N_k>n_k)$, we have that $P(T-t_n>\log z)\leq \exp(-\rho\log z/2)$.
Hence, it holds that $\log\left[\frac{h_2'(\t,\y)}{h_2(\t,\y)}\right]=o((\log z)^2)$ with probability at least $1-c_0z^{-\rho/2}$ for some positive constant $c_0$. Thus, we have shown \eqref{eqn:bias}. 
}

{Correspondingly, applying \eqref{eqn:bias} to the joint density of $\t$ and $\y$ in each sub-window and taking $z=T^{2/\rho}$, we have $\sup_{\y_m}\left[\log f(\t_m,\y_m|\btheta)-\log f_0(\t_m,\y_m|\btheta)\right]=o((\log T)^2)$ with probability at least $1-2c_0T^{-1}$ for some positive constant $c_0$.  Jointly for all sub-windows, we have 
\begin{equation}\label{likelihoodbias}
 \log\left[\prod_{m=1}^M \sum_{\y_m}f(\t_m,\y_m|\btheta)\right]= \log\left[\prod_{m=1}^M\sum_{\y_m} f_0(\t_m,\y_m|\btheta)\right]+o(M(\log T)^2),
\end{equation}
with probability at least $1-2c_0M T^{-1}$. 
The discussions for the bivariate case follows an almost identical argument as in the univariate case, and we omit deriving \eqref{likelihoodbias} for the bivariate case in this proof.
}

{
Noting $\ell_0(\btheta;\t,\x)=\sum_m\log\left[\sum_{\y_m} f_0(\t_m,\y_m|\btheta)\right]$ and $\ell_s^c(\btheta;\t,\x)=\sum_m\log\left[\sum_{\y_m} f(\t_m,\y_m|\btheta)\right]$, and we have
\begin{equation}\label{ellbias}
    \ell_0(\btheta;\t,\x)=\ell_s^c(\btheta;\t,\x)+o(M(\log T)^2/T),
\end{equation}
with probability at least $1-2c_0MT^{-1}$. Here, $\ell_0(\btheta;\t,\x)$ is the true composite likelihood and $\ell_s^c(\btheta;\t,\x)$ is the approximated composite likelihood. 
}

{
As assumed in Theorem 3, the eigenvalues of $-\mathbb{E}\left\{\nabla_{\btheta}^2\ell_0(\btheta;\t,\x)\right\}$ and $-\mathbb{E}\left\{\nabla_{\btheta}^2\ell_s^c(\btheta;\t,\x)\right\}$ are lower bounded by $\nu_1>0$ for $\btheta\in\mathcal{B}_{r_0}(\btheta_0)$, where $\mathcal{B}_{r_0}(\btheta_0)$ denotes the Frobenius-norm ball around $\btheta_0$ with radius $r_0$ and $r_0$ is a positive constant. By Taylor's expansion, we get that
\begin{eqnarray*}
&&\mathbb{E}\left\{\ell_0(\t,\btheta_s^\ast)\right\}-\mathbb{E}\left\{\ell_0(\t,\btheta_0)\right\}\\
&=&\left\langle\mathbb{E}\left\{\nabla_{\btheta}\ell_0(\btheta;\t,\x)\right\}|_{\btheta=\btheta_0},\btheta_s^\ast-\btheta_0 \right\rangle+(\btheta_s^\ast-\btheta_0)^\top\mathbb{E}\left\{\nabla_{\btheta}^2\ell_0(\btheta;\t,\x)\right\}|_{\btheta=\btheta_1}(\btheta_s^\ast-\btheta_0),
\end{eqnarray*}
where $\btheta_1$ is between $\btheta_s^\ast$ and $\btheta_0$. Since $\mathbb{E}\left\{\nabla_{\btheta}\ell_0(\btheta;\t,\x)\right\}|_{\btheta=\btheta_0}=\0$, we have that
\begin{equation*}
\mathbb{E}\left\{\ell_0(\t,\btheta_0)\right\}-\mathbb{E}\left\{\ell_0(\t,\btheta_s^\ast)\right\}\geq \nu_1\|\btheta_s^\ast-\btheta_0\|_2^2.
\end{equation*}
Similarly, by Conditions (2.1), we can show that $\mathbb{E}\left\{\ell(\t,\btheta_s^\ast)\right\}-\mathbb{E}\left\{\ell(\t,\btheta_0)\right\}\geq \nu_1\|\btheta_s^\ast-\btheta_0\|_2^2$.
Combining the two above equations, it arrives at
\begin{equation*}
    \mathbb{E}\left\{\ell(\t,\btheta_s^\ast)\right\}-\mathbb{E}\left\{\ell(\t,\btheta_0)\right\}+\mathbb{E}\left\{\ell_0(\t,\btheta_0)\right\}-\mathbb{E}\left\{\ell_0(\t,\btheta_s^\ast)\right\}\geq 2\nu_1\|\btheta_s^\ast-\btheta_0\|_2^2.
\end{equation*}
By \eqref{ellbias}, we have $\mathbb{E}\left\{\ell(\t,\btheta_s^\ast)\right\}-\mathbb{E}\left\{\ell(\t,\btheta_0)\right\}+\mathbb{E}\left\{\ell_0(\t,\btheta_0)\right\}-\mathbb{E}\left\{\ell_0(\t,\btheta_s^\ast)\right\}=o(M(\log T)^2/T)$. Assuming that $M=O(T^{2/5})$, we have $M(\log T)^2/T=o(T^{-1/2})$ and hence $\|\btheta_s^\ast-\btheta_0\|_2=o(T^{-1/2})$. We arrive at the desired result in Step 2.
}

\section*{B. Computational details and results}

\subsection*{B.1. Maximization in the M-step}
At the $p$-th composite likelihood EM iteration, let $\pi_{lh}(\btheta_{p-1})$, $t_l\in[ms-s,ms)$ denote the estimated $P(Y_l=h|\t_m,\x_m,\btheta_{p-1})$ in the composite likelihood E-step. With some algebra, we can show that
$$
\hat n_k=\sum_{l=2}^nI(|x_l-x_{l-1}|>0)+\sum_{l=2}^nI(|x_l-x_{l-1}|=0)\pi_{l1}(\btheta_{p-1})+1,
$$
$$
\hat n_{k1}=\sum_{l=2}^n I(x_{l-1}-x_l=1)+I(x_n=1)+\sum_{l=2}^nx_lx_{l-1}\pi_{l1}(\btheta_{p-1}),
$$
$$
\hat n_{k0}=\sum_{l=2}^n I(x_l-x_{l-1}=1)+I(x_n=0)+\sum_{l=2}^n(1-x_l)(1-x_{l-1})\pi_{l1}(\btheta_{p-1}),
$$
where $\hat n_k$ is the estimated total number of segments, $\hat n_{k1}$ is the estimated total number of original post segments and $\hat n_{k0}$ is the estimated total number of repost segments. The calculations are straightforward and we omit details here.

In the M-step, the $\alpha$, $\gamma$, $\mu_1$, $\mu_0$ in $\btheta_{p}$ can be updated using
$$
\hat\alpha={\sum_{l=1}^n \pi_{l1}(\btheta_{p-1})x_l\over \sum_{l=1}^n \pi_{l1}(\btheta_{p-1})},\quad \hat\gamma=\frac{\hat n_k}{\sum_{l=1}^n \pi_{l1}(\btheta_{p-1})}-1,
$$
$$
\hat\mu_1=\frac{\sum_{l=1}^nx_l}{\hat n_{k1}}-1,\quad \hat\mu_0=\frac{n-\sum_{l=1}^nx_l}{\hat n_{k0}}-1.
$$

To update $\bbeta$, we need to solve
\begin{eqnarray*}
\sum_{l=1}^n\pi_{l1}(\btheta_{p-1}){\lambda^{(1)}(t_l;\bbeta)\over \lambda(t_l;\bbeta)}-\sum_{l=1}^n\pi_{l1}(\btheta_{p-1})\int_{t_{l-1}}^{t_l} \lambda^{(1)}(t;\bbeta)\dd t-\int_{t_n}^T \lambda^{(1)}(t;\bbeta)\dd t={\bf 0},
\end{eqnarray*}
where $\lambda^{(1)}(t;\bbeta)$ is the first-order derivative of $\lambda(t;\bbeta)$ with respect to $\bbeta$. This can be solved using standard numerical methods.
Furthermore, as the offspring gap times follow exponential distributions as in (\ref{expgap}), $\rho_1$ and $\rho_0$ can be updated using
$$
\hat\rho_1={\sum_{l=1}^n \pi_{l0}(\btheta_{p-1})x_l\over \sum_{l=1}^n \pi_{l0}(\btheta_{p-1})x_ld_l},\quad\hat\rho_0={\sum_{l=1}^n \pi_{l0}(\btheta_{p-1})(1-x_l)\over \sum_{l=1}^n \pi_{l0}(\btheta_{p-1})(1-x_l)d_l}.
$$

\subsection*{B.2. Conditional expectation calculation}
\label{sec::condest}
In the E-step of the CLEM algorithm, we need to calculate the conditional distribution, i.e.,
\begin{eqnarray}\label{eq:condP}
P_{\btheta}(Y_l=h|\t_m,\x_m,\btheta)={\sum_{\y_m|y_l=h} f(\t_m,\x_m,\y_m|\btheta)\over \sum_{\y_m} f(\t_m,\x_m,\y_m|\btheta)},\quad h=0,1.
\end{eqnarray}
If we can identify several parent events a priori, the computing cost in the E-step can be reduced.
To simplify notation in this section, we suppress the notation $m$ that is used to index the sub-window. 
Suppose we identify $\tilde k$ parent events, denoted as $t_{p_i}$, $i\in[\tilde k]$ where $1=p_1<\cdots<p_{\tilde k}$.
We may divide $\t$, $\x$ and $\y$ into $\tilde k$ non-overlapping segments $\t_1,\ldots,\t_{\tilde k}$, $\x_1,\ldots,\x_{\tilde k}$ and $\y_1,\ldots,\y_{\tilde k}$ respectively, where $\t_i=(t_{p_{i}},\ldots,t_{p_{i+1}-1})$, $\x_i=(x_{p_{i}},\ldots,x_{p_{i+1}-1})$, $\y_i=(y_{p_{i}},\ldots,y_{p_{i+1}-1})$, $i\in[\tilde k-1]$. Write $\t_{\tilde k}=(t_{p_{\tilde k}},\cdots,t_n)$, $\x_{\tilde k}=(x_{p_{\tilde k}},\cdots,x_n)$ and $\y_{\tilde k}=(y_{p_{\tilde k}},\cdots,y_n)$. With some straightforward algebra, it can then be shown that
\begin{equation}\label{condE}
f(\t,\x,\y|\btheta)=\prod_{i=1}^{\tilde k} f_i(\t_i,\x_i,\y_i|\btheta),
\end{equation}
where $f_i(\t_i,\x_i,\y_i|\btheta)$'s are defined as follows.
Suppose there are $\tau_i$ episodes prior to $t_{p_i}$, $i\in[\tilde k]$. We have
\begin{eqnarray*}
f_1(\t_1,\x_1,\y_1|\btheta)&=&\prod_{l=1}^{p_2-1}f_{l}(d_l)\times\prod_{l=1}^{p_2-1}\alpha^{I(y_l=1,x_l=1)}(1-\alpha)^{I(y_l=1,x_l=0)}\\
&&\times\prod_{k=1}^{\tau_2} \frac{\gamma^{n_k-1}e^{-\gamma}}{(n_k-1)!}\times\prod_{k=1}^{\tau_2} \prod_{j=1}^{n_k}\frac{(\mu_1^{l_{k_j}-1}e^{-\mu_1})^{I(z_{k_j}=1)}(\mu_0^{l_{k_j}-1}e^{-\mu_0})^{I(z_{k_j}=0)}}{(l_{k_j}-1)!},
\end{eqnarray*}
Furthermore, for $1< i <\tilde k$,
\begin{eqnarray*}
f_i(\t_i,\x_i,\y_i|\btheta)&=&f_{p_i,1}(d_{p_i};\btheta)\prod_{l=p_i+1}^{p_{i+1}-1}f_{l}(d_l)\times\prod_{l=p_i+1}^{p_{i+1}-1}\alpha^{I(y_l=1,x_l=1)}(1-\alpha)^{I(y_l=1,x_l=0)}\\
&&\times\prod_{k=\tau_i+1}^{\tau_{i+1}} \frac{\gamma^{n_k-1}e^{-\gamma}}{(n_k-1)!}\times\prod_{k=\tau_i+1}^{\tau_{i+1}} \prod_{j=1}^{n_k}\frac{(\mu_1^{l_{k_j}-1}e^{-\mu_1})^{I(z_{k_j}=1)}(\mu_0^{l_{k_j}-1}e^{-\mu_0})^{I(z_{k_j}=0)}}{(l_{k_j}-1)!},
\end{eqnarray*}
and $f_{\tilde k}(\t_{\tilde k},\x_{\tilde k},\y_{\tilde k}|\btheta)$ can be derived analogously. 
It can be easily seen that $\sum_{\y} f(\t,\x,\y|\btheta)=\prod_{i=1}^{\tilde k}\sum_{\y_i} f_i(\t_i,\x_i,\y_i|\btheta))$ given $y_{p_1}=\cdots=y_{p_{\tilde k}}=1$, and calculating $P_{\btheta}(Y_l=m|\t,\x)$ can be simplified as
\begin{eqnarray}\label{eq:condP2}
{P_{\btheta}(Y_l=m|\t,\x)={\sum_{\y_i|y_l=m} f_i(\t_i,\x_i,\y_i|\btheta)\over \sum_{\y_i} f_i(\t_i,\x_i,\y_i|\btheta)}},\hbox{   }y_l\in\y_i.
\end{eqnarray}
This is much easier to calculate because the summations in (\ref{eq:condP2}) are over subsets of $\y$ rather than $\y$. 

\subsection*{B.3. Goodness of fit}
\label{sec::good}
In this section, we propose a goodness-of-fit procedure that compares the empirical gap time distribution to that calculated from realizations simulated from the fitted model. 

The gap time distribution function from the observed data, denoted as $\hat F(v)$, is calculated as:
\[
\hat F(v)=\frac{1}{n}\sum_{l=1}^nI(d_l<v),
\]
where $d_l=t_l-t_{l-1}$, $l\in[n]$.
We can calculate the distribution functions denoted as $\hat F^{(i)}(v)$, $i\in[w]$, from $w$ independent realizations in $[0,T]$ from the fitted model.
Define
$$
\bar F(v)={1\over w}\sum_{i=1}^w \hat F^{(i)}(v), \quad U(v)=\max\{\hat F^{(i)}(v)\}, \quad L(v)=\min\{\hat F^{(i)}(v)\}.
$$
To evaluate the goodness of fit, we plot $\hat F(v)$ against $\bar F(v)$ along with the upper and lower simulation envelopes $U(v)$ and $L(v)$.
If the fitted model is compatible with the observed data, the plot of $\hat F(v)$ against $\bar F(v)$ should be roughly linear and contained in the simulation envelopes.

We may wish to further investigate the gap time distributions for offspring original posts, offspring reposts and parent posts.
Denote the gap times for the offspring original posts and reposts by $E_1$ and $E_0$, respectively. Define $F_i(v)=P(E_i<v)$, $i=0,1$.
From the estimated model, both $F_1(v)$ and $F_0(v)$ can be easily calculated. As we assume an exponential distribution for $E_1$, then $F_1(v)=1-\exp(-\hat\rho_1 v)$. Furthermore, we can estimate $F_1(v)$ with
\begin{equation}\label{F1}
\hat F_1(v)=\frac{\sum_{l=1}^n\pi_{l0}(\hat\btheta_{M,s})I(x_l=1)I(d_l<v)}{\sum_{l=1}^n\pi_{l0}(\hat\btheta_{M,s})I(x_l=1)},
\end{equation}
where $\pi_{l0}(\hat\btheta_{M,s})=P(Y_l=0|\t_m,\x_m,\hat\btheta_{M,s})$, $t_l\in[ms-s,s)$ and $\hat\btheta_{M,s}$ is the estimate of $\btheta$ from the proposed CLEM algorithm. 
To assess the goodness of fit, we can compare $\hat F_1(v)$ to $F_1(v)$ over a range of different $v$ values. The goodness of fit for offspring reposts can be evaluated similarly by comparing $\hat F_0(v)$ against $F_0(v)$, where $\hat F_0(v)$ can be calculated analogous to (\ref{F1}).

To assess the goodness of fit for the parent event gap time distribution, we use the following result. Assume we observe event time locations at $0=u_0<u_1<u_2<\ldots<u_n<T$. Let the gap times $u_l-u_{l-1}$, $l=1,\ldots,n$, follow the density function in (\ref{pargap}).
Define
$$
\Lambda(u_l)=\int_{u_{l-1}}^{u_l}\lambda(u)du, \quad l=1,\ldots,n.
$$
Then, $\Lambda(u_l)$'s follow an exponential distribution with the unit rate.
This is a special case of the time change theorem from \cite{Meyer1971}. Thus, we can rescale the inhomogeneous parent gap times as random variables from an exponential distribution with the unit rate. Define
$$
\hat F_2(v)=\frac{\sum_{l=1}^n\pi_{l1}(\hat\btheta_{M,s})I\left[\int_{t_{l-1}}^{t_l}\lambda(t,\hat\bbeta)dt<v\right]}{\sum_{l=1}^n\pi_{l1}(\hat\btheta_{M,s})}.
$$
If the estimated parent gap time distribution fits the observed pattern well, $\hat F_2(v)$ should be close to $F_2(v)=P(E_2<v)$, where $E_2\sim\exp(1)$.

\subsection*{B.4. Fitting the bivariate Hawkes process in \eqref{hawkes}}
\label{sec:B4}
{For both models, we set $\phi(x)=\max\left\{x,0 \right\}$ and approximated the transfer functions using cubic B-splines. For the nonstationary model, we modeled the background intensity functions using cyclic cubic B-splines as follows:
$$
\nu_i(t)=\sum_{k=1}^q\beta_k B_k(t-\floor{t}), \;\; i=1,2,
$$
where $B_k(\cdot)$ and $\beta_k$, $k=1,\ldots,q$, are $q$ cyclic B-spline basis functions defined on $[0,1]$ and the associated coefficients, respectively. }
The background intensity functions $\mu_1(t)$ and $\mu_2(t)$ are modeled using cyclic cubic B-splines, defined on [0,1] with nine internal knots. The transfer functions are modeled using cubic B-splines defined on [0,0.03] with two internal knots. The range $[0,0.03]$ for the excitatory functions equals to approximately 45 minutes. We have also tried bigger and smaller ranges, and the results remained similar. Parameter estimation is carried out by minimizing a squared loss function \citep{cai2020latent}.

\subsection*{B.5. Additional simulation results} 
{
In this section, we carry out simulations under a misspecified setting where the event locations are generated from a bivariate Hawkes process, one of the most popular cluster point process models.
Recall from Section 2, the intensity functions of a bivariate Hawkes process take the form
\begin{equation*}
\lambda_i(t)=\nu_i(t)+\int_0^t\omega_{ii}(t-s) N_i(\dd s)+\int_0^t\omega_{ij}(t-s) N_j(\dd s),
\end{equation*}
where $\nu_i(t)>0$ is the background intensity for the $i$th point process, and $\omega_{ii}(\cdot)$ and $\omega_{ij}(\cdot)$ are some transfer functions, for $i,j=1,2$ and $i\ne j$. 
We set $\nu_1(t)=4+4\sin(2\pi t)$, $\nu_2(t)=4+4\sin(2\pi t)$ and 
\begin{equation*}
\begin{aligned}
 \omega_{11}(t)=2\exp(-10 t), &\quad \omega_{12}(t)=\exp(-30 t),\\
 \omega_{21}(t)=0, &\quad \omega_{22}(t)=2\exp(-20 t).
\end{aligned}
\end{equation*}
We simulate data from the above bivariate Hawkes process with $T=100$ and fit our proposed model and the Hawkes process model. For the Hawkes process model, we approximate the transfer functions using cubic B-splines and the background intensity functions using cyclic cubic B-splines (see details in Section B.4 of the supplementary material).
We then adopt the goodness-of-fit procedure in Section B.3 to evaluate the performance of both fitted models. 
\begin{figure}[!t]
\centering
\includegraphics[trim=2cm 0 1cm 0, scale=0.35]{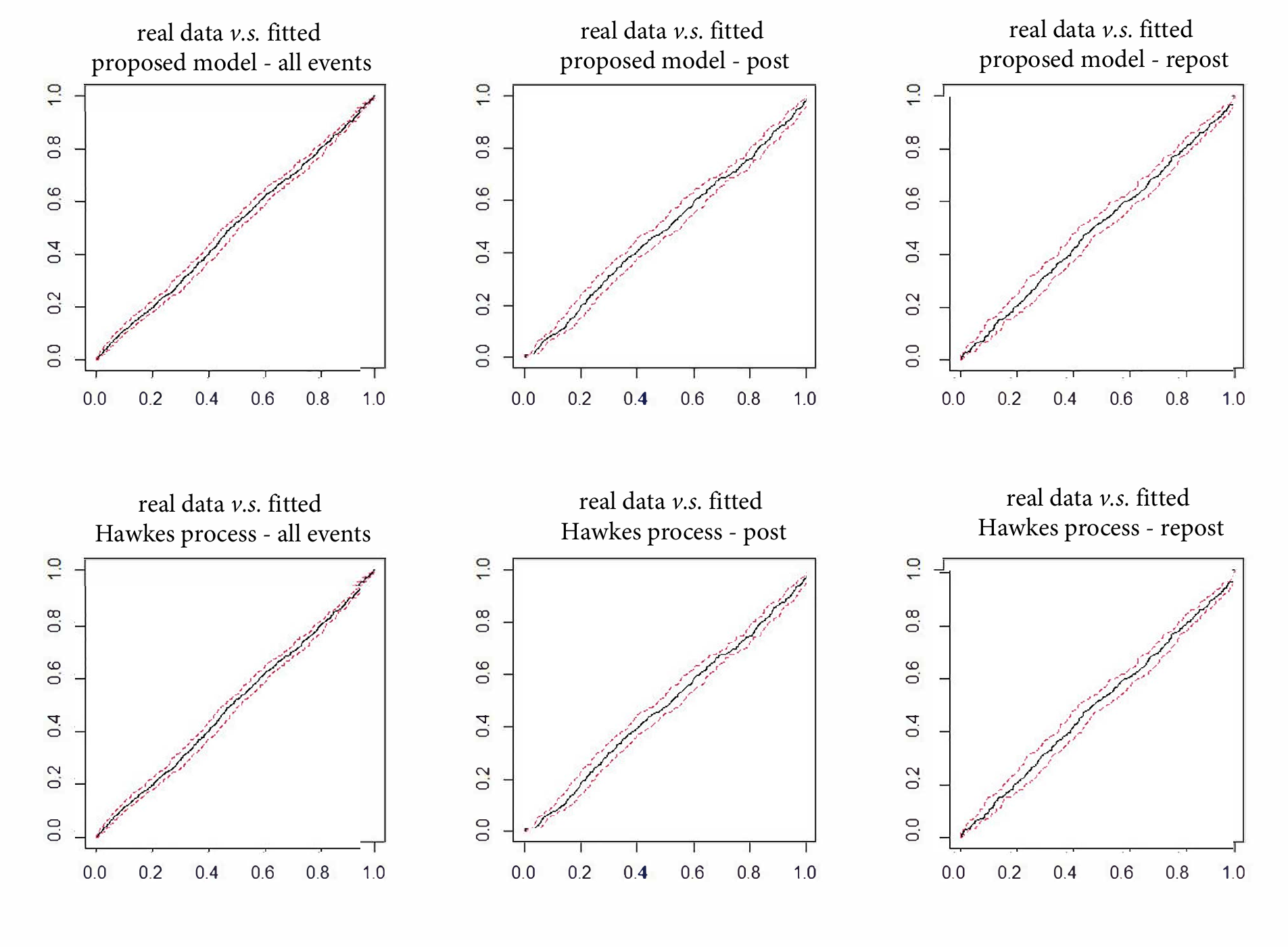}
\includegraphics[trim=0 -1.5cm 0 0, scale=0.325]{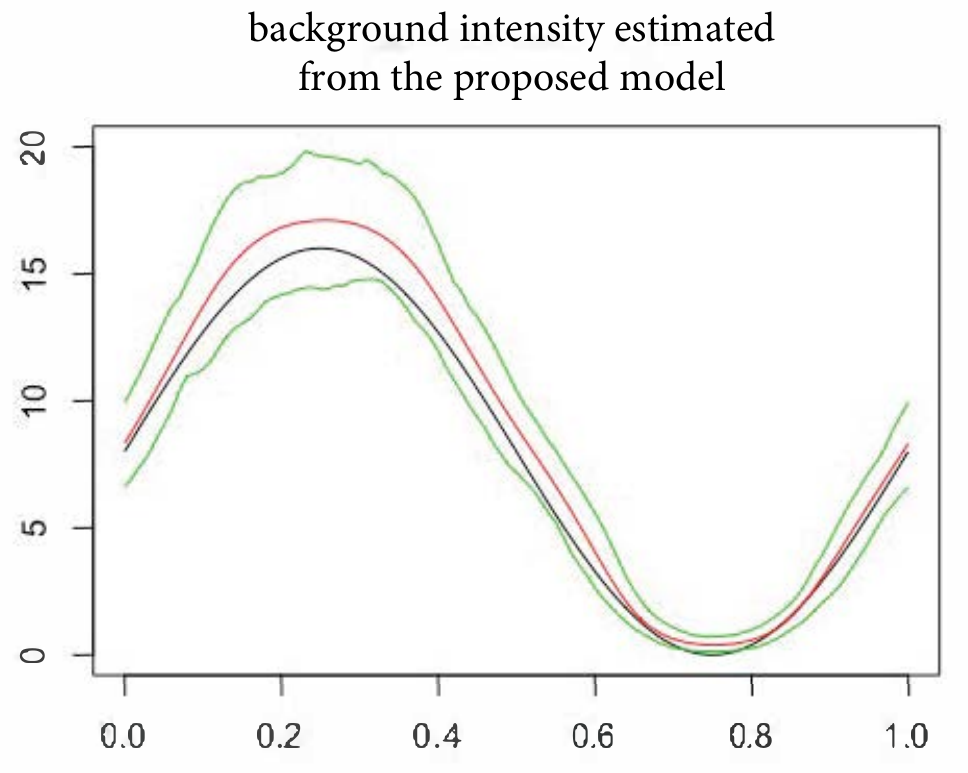}
\caption{{Goodness-of-fit plots (left panel) and the estimated background intensity (right panel) from the proposed model. In right panel, the black curve shows the true intensity $\nu_1(t)$, the curve line shows the mean intensity of 100 replications and the green curves mark the 95$\%$ interval.}}
\label{fig3}
\end{figure}
Figure~\ref{fig3} presents the goodness-of-fit plots (for all events, post events only and repost events only) and the estimated background intensity, averaged over 100 data replicates. It is seen that our proposed model fits the data well even when it is misspecified, by noting that the empirical gap times against estimated gap times is roughly linear and contained in the simulation envelopes. 
Moreover, the estimated hazard function for a parent event in our proposed model gives a reasonable approximation to the background intensity in the Hawkes process. 
}

\section*{C. Additional plots from Section \ref{sec:real}}
We present the goodness-of-fit plots model fitted for January 2017, the first month of Trump's presidency.
\begin{figure}[!t]
\centering
\makebox{\includegraphics[width=1\textwidth]{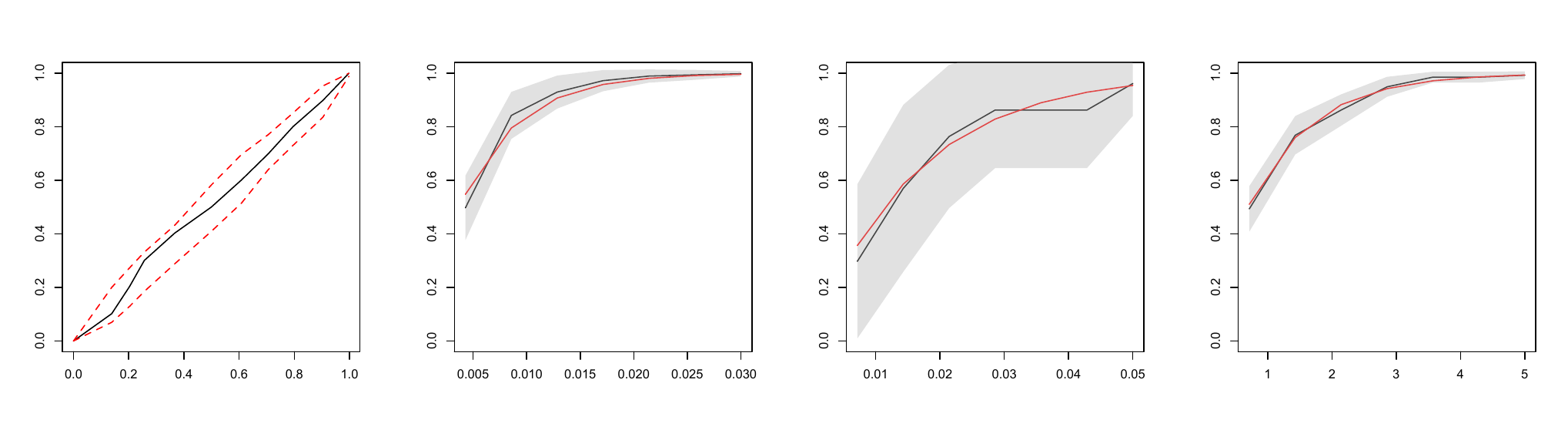}}
\caption{Goodness-of-fit plots of the proposed model for January 2017. From left to right are the envelop plot with the upper and lower envelopes marked in red dashed lines, goodness-of-fit plots for the offspring original post, offspring repost and parent inter-event distances. The red solid lines in the last three plots are calculated from cdfs of the fitted exponential distributions. The grey bands are the 95\% confidence intervals.}
\label{fig9}
\end{figure}
For the envelope plot, we simulate 99 realizations from the fitted model. We can see that the $\hat F(v)$ against $\bar F(v)$ line is roughly linear and contained in the simulation envelope. This suggests that the simulated gap times match the observed ones. Furthermore, we compare $\hat F_i(v)$ against $F_i(v)$, $i=0,1,2$, in the last three plots of Figure~\ref{fig9}. For the confidence intervals, the standard error of $\hat F_i(v)$ for a given $v$ is approximated by assuming that distributions of gap times $D_l$, $l\in[N]$ are independent in the calculation. We can see that the estimated gap time distributions (i.e., $\hat F_i(v)$'s) appear to be in close agreement with their theoretical counterparts (i.e., $F_i(v)$'s) from the fitted model.

\begin{figure}[!t]
\centering
\makebox{\includegraphics[trim=5mm 5mm 0 5mm, scale=0.275]{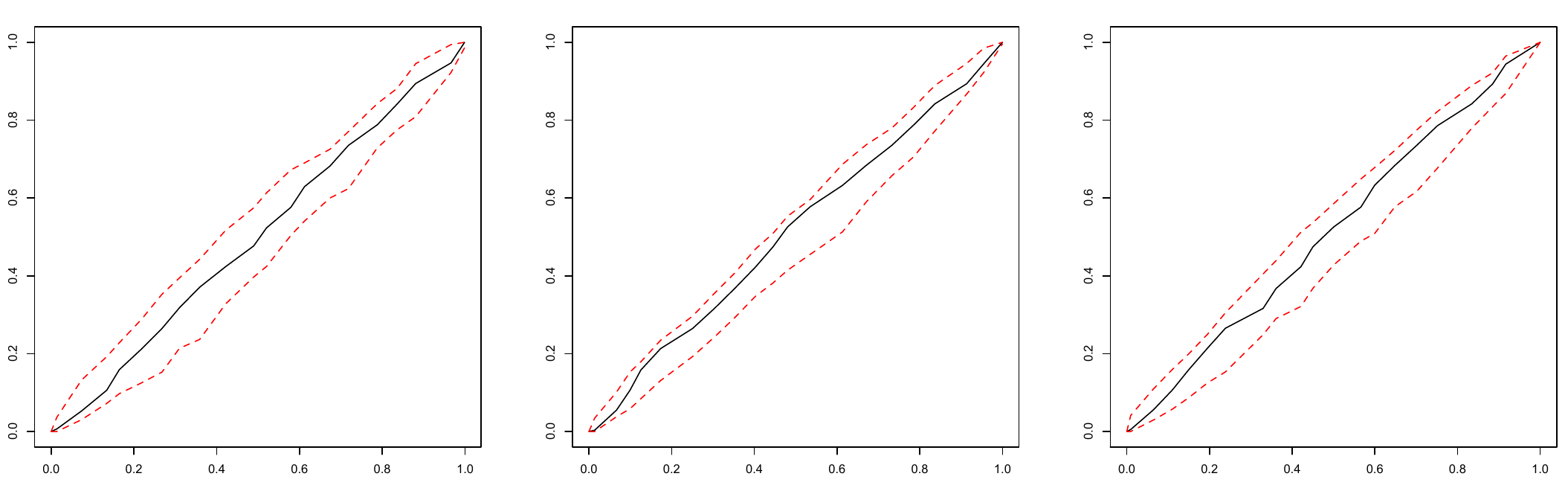}}
\makebox{\includegraphics[trim=5mm 0mm 0 0mm,scale=0.35]{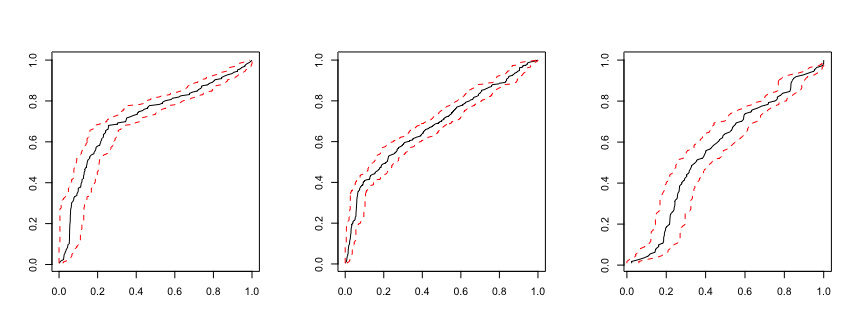}}
\caption{The goodness-of-fit envelope plots for three randomly selected users using our proposed method (top panel) and the bivariate Hawkes process (bottom panel). Upper and lower envelopes are marked in red dashed lines.}
\label{fig8}
\end{figure}

\begin{figure}[!t]
\centering
\makebox{\includegraphics[trim=1cm 0 0 1.5cm, scale=0.375]{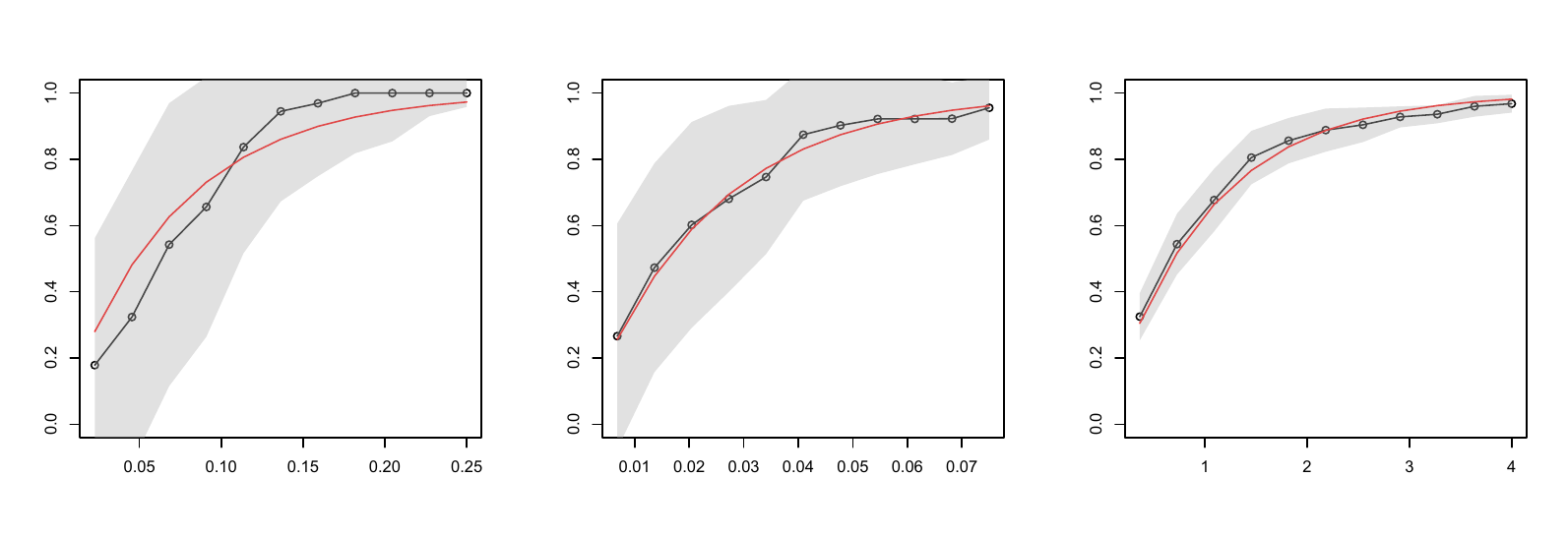}}
\makebox{\includegraphics[trim=1cm 0 0 1.5cm, scale=0.375]{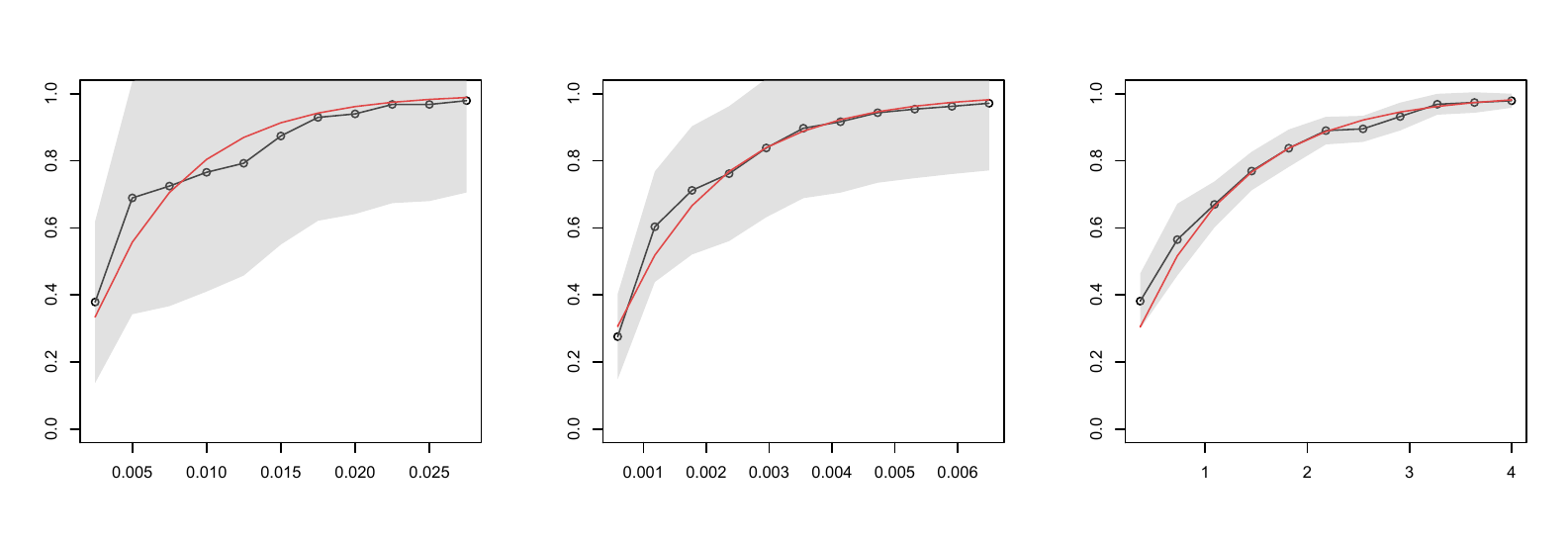}}
\makebox{\includegraphics[trim=1cm 0 0 1.5cm, scale=0.375]{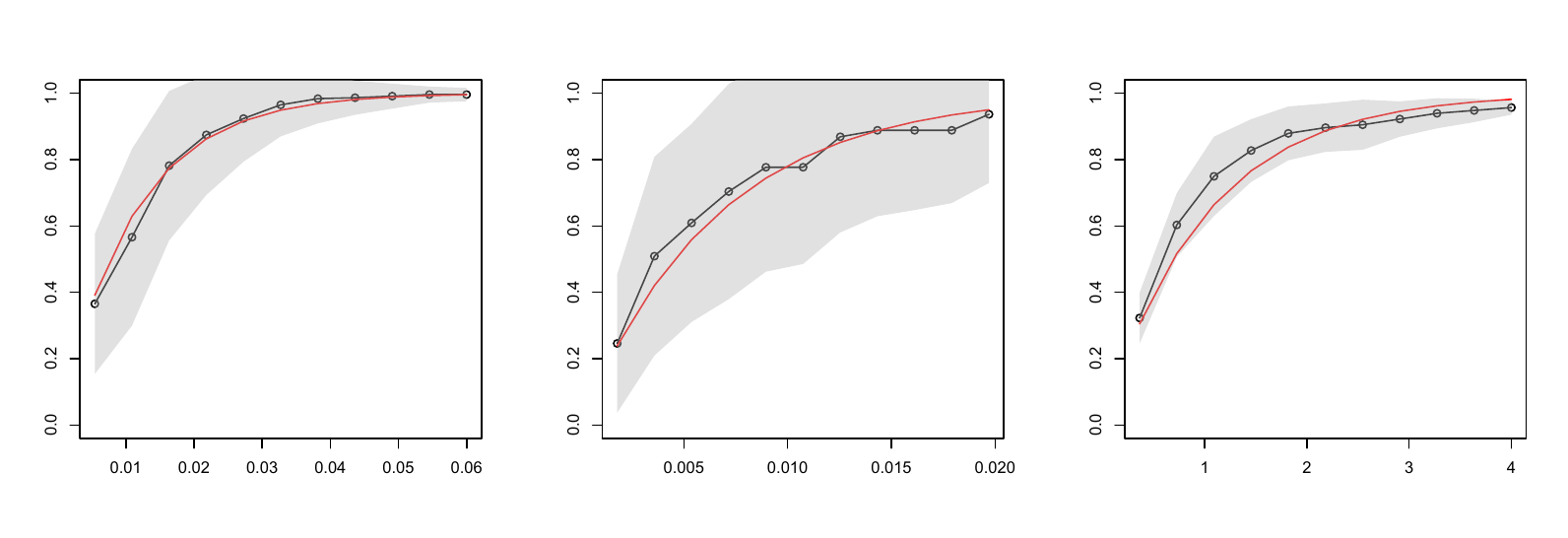}}
\caption{The goodness-of-fit plots for the original offspring post (left plot), offspring repost (middle plot) and parent (right plot) inter-event distances of user 1 (top panel), user 2 (middle panel) and user 3 (bottom panel). Red solid lines are calculated from the cdf of exponential distributions. The grey bands are the 95\% confidence intervals.}
\label{fig4}
\end{figure}

We present the goodness-of-fit plots for three randomly selected users in Figures \ref{fig8} and \ref{fig4}.

\end{document}